\documentclass[aps,reprint,showpacs,amsmath,amssymb,prc,floatfix,nofootinbib]{revtex4-1}
\usepackage{color}
\usepackage{graphicx}
\usepackage{dcolumn}
\usepackage{bm}
\usepackage{rotating}
\usepackage{float}
\usepackage{xcolor}
\usepackage{natbib}
\usepackage{color}
\usepackage{times}
\usepackage{algorithm2e}
\usepackage{inputenc}
\usepackage{multirow}
\usepackage[colorlinks,citecolor=blue,linkcolor=red,anchorcolor=blue,filecolor=blue,urlcolor=blue]{hyperref}
\usepackage{comment}
\begin{document}
\title{Incompressibility and Symmetry Energy of Neutron Star }
\author{Ankit Kumar$^{1,2}$}
\email{ankit.k@iopb.res.in}
\author{H. C. Das$^{1,2}$}
\email{harish.d@iopb.res.in}
\author{S. K. Patra$^{1,2}$}
\email{patra@iopb.res.in}
\affiliation{\it $^{1}$Institute of Physics, Sachivalaya Marg, Bhubaneswar 751005, India}
\affiliation{\it $^{2}$Homi Bhabha National Institute, Training School Complex,
Anushakti Nagar, Mumbai 400094, India}
\date{\today}
\begin{abstract}
We trace a systematic and consistent method to precisely numerate the magnitude range for various structural and isospin compositional properties of the neutron star. Incompressibility, symmetry energy, slope parameter and curvature of a neutron star are investigated using the relativistic energy density functional within the framework of coherent density fluctuation model. The analytical expression for the energy density functional of the neutron star matter is motivated from the Br$\ddot{u}$ckner functional and acquired by the polynomial fitting of the saturation curves for three different relativistic mean-field parameter sets (NL3, G3 and IU-FSU). The modified functional is folded with the neutron star's density-dependent weight function to calculate the numerical values for incompressibility and symmetry energy using the coherent density fluctuation model. NL3 parameter set, being the stiffest equation of state, endue us with a higher magnitude of all the properties compared to the other two parameter sets.    
\end{abstract}
\maketitle
Neutron stars, being the extremely and suprisingly dense objects in the universe, are astounding to both astrophysicists and nuclear physicists in their respective manner. Astronuclear physicists perceive the neutron stars as the enormously large nucleus bound by the gravity and amazed by its short-range strong nuclear interactions as well as its enormous gravitational and electromagnetic interactions. An entire and appropriate understanding of the properties of neutron stars demand a thorough knowledge of both the astrophysics and the nuclear physics, which are mostly incoherent. An inevitable conformability exits between the astral properties (i.e. gravitational potential, central temperature, angular velocity, magnetic poles) and the nuclear properties (i.e. baryon density, neutrino emissivity, isospin asymmetry, superconductivity) of the neutron stars \cite{Lattimer536}. However, there are some properties of the neutron stars like pressure, incompressibility and symmetry energy, which are tremendously significant for the experimental and theoretical explanations of its nontrivial behaviour.

The consequences of ``Symmetry" are inalienable to many important aspects of the modern physics. The theoretical modelling to study the analytical and structural behaviour of highly asymmetric dense nuclear matter depends significantly on the symmetry energy \cite{LATTIMER2014276}. Also, the parameters derived by expanding the symmetry energy around saturation density (slope and curvature parameter) controls the core-crust transition density, transition pressure and the cooling rate of the neutron stars \cite{PhysRevC.94.052801, PhysRevC.100.055802}. The quantitative information on the slope and curvature parameter of the neutron stars can be applied to constraint the equation of state (EoS) obtained using different parameter sets of the vast relativistic framework. The particle fraction in the core of the neutron stars, which is a critical quantity to the cooling of neutron stars, is also controlled by the domain of symmetry energy and slope parameter \cite{PhysRevC.103.034330}. Since, symmetry energy can not be measured directly, so, the elucidation of experimentally available data requires a substantial and consistent theoretical model for nuclear matter at very high densities \cite{ZHANG2008145, Hagel:2014wja, PhysRevC.85.064618}. The precise range of the symmetry energy for the isospin asymmetric nuclear matter and the neutron stars has been a questionable issue among the researchers incessantly. Along with the theoretical calculations, laboratory nuclear experiments (i.e. giant dipole resonance, heavy-ion collisions, neutron skin) and abridged results of astrophysical observations (i.e. mass-radius profile, dimensional tidal deformability) has also put some constraints on the domain of symmetry energy and its derivatives \cite{Kumar_2020, ZHANG2013234, PhysRevC.77.061304, PhysRevLett.102.122701}. Another important quantity which controls the equation of state (EoS) of any dense matter system is the incompressibility.

Dependence of incompressibility on the isospin asymmetry parameter of the nuclear system restricts the stiffness of the EoS which indirectly influences the maximum mass and radius of the neutron star e.g. maximally compressible dense matter handles the upper limit on the maximum mass of the neutron star \cite{PhysRevC.63.015802}. The quantitative dimension of the incompressibility coefficient also helps in the determination of angular velocity and the evolution of cooling stage of the rapidly rotating neutron stars \cite{1994ApJ...436..257H}. The collective motion of the nucleons inside a nucleus produce the nuclear giant resonances, which is a global feature of the finite nuclei \cite{RevModPhys.55.287, BERTRAND1981129}. The iso-scalar giant monopole resonance (ISGMR) is the most common collective oscillation with both the protons and neutrons being in same phase. This ISGMR  is the breathing  mode oscillation related to the incompressibility $K_A$ of the finite nucleus of mass A. Thus, the nuclear incompressibility $K_A$ is a vital quantity to understand the various modes of oscillations of the finite nucleus \cite{PhysRevC.69.034315, Buenerd}. Similarly the other modes of collective oscillations, such as iso-vector giant dipole resonance (IVGDR), iso-scalar/iso-vector giant quadrupole resonances (ISGQR/IVGQR) are also governed by the nuclear incompressibility and the other symmetry related parameters of the finite nucleus. The various collective resonances decide the internal structures of the finite nucleus, for example, the IVGDR tells the shape of the nucleus and the ISGMR gives information about the compression or expansion capacity of the nucleus.  The calculations of the monopole and quadrupole excitation energies and their relations with incompressibility using various sum rule approaches are demonstrated in \cite{PhysRevC.72.014304, Patra_2002}. For a particular model, either a relativistic or a non-relativistic, to be consistent with the constraints set by the terrestrial or astrophysical experiments, the nuclear incompressibility, symmetry energy and other related quantities are very much essential. Neutron star (NS), being a huge nucleus hypothetically, with mass number $A\sim 10^{57}$, it must posses all the natural properties of practical finite nuclei, like all type of collective oscillations. So, the incompressibility $K^{star}$, symmetry energy $S^{star}$ and its higher derivatives like $L^{star}_{sym}$, $K^{star}_{sym}$ etc. are quite informative and necessary to explore the structure of the neutron star. In the present paper, our aim is to bring the attention of a new approach, i.e., Coherent Density Fluctuation Model (CDFM) for the evaluation of these quantities of neutron stars. This model is applicable for neutron star, as it has a finite surface like that of a standard nucleus. Here, we first time provide an approach to calculate both of these parameters (symmetry energy and incompressibility) for a neutron star matter system in a consistent and accurate manner.   

\begin{figure}
\centering
\includegraphics[width=1\columnwidth]{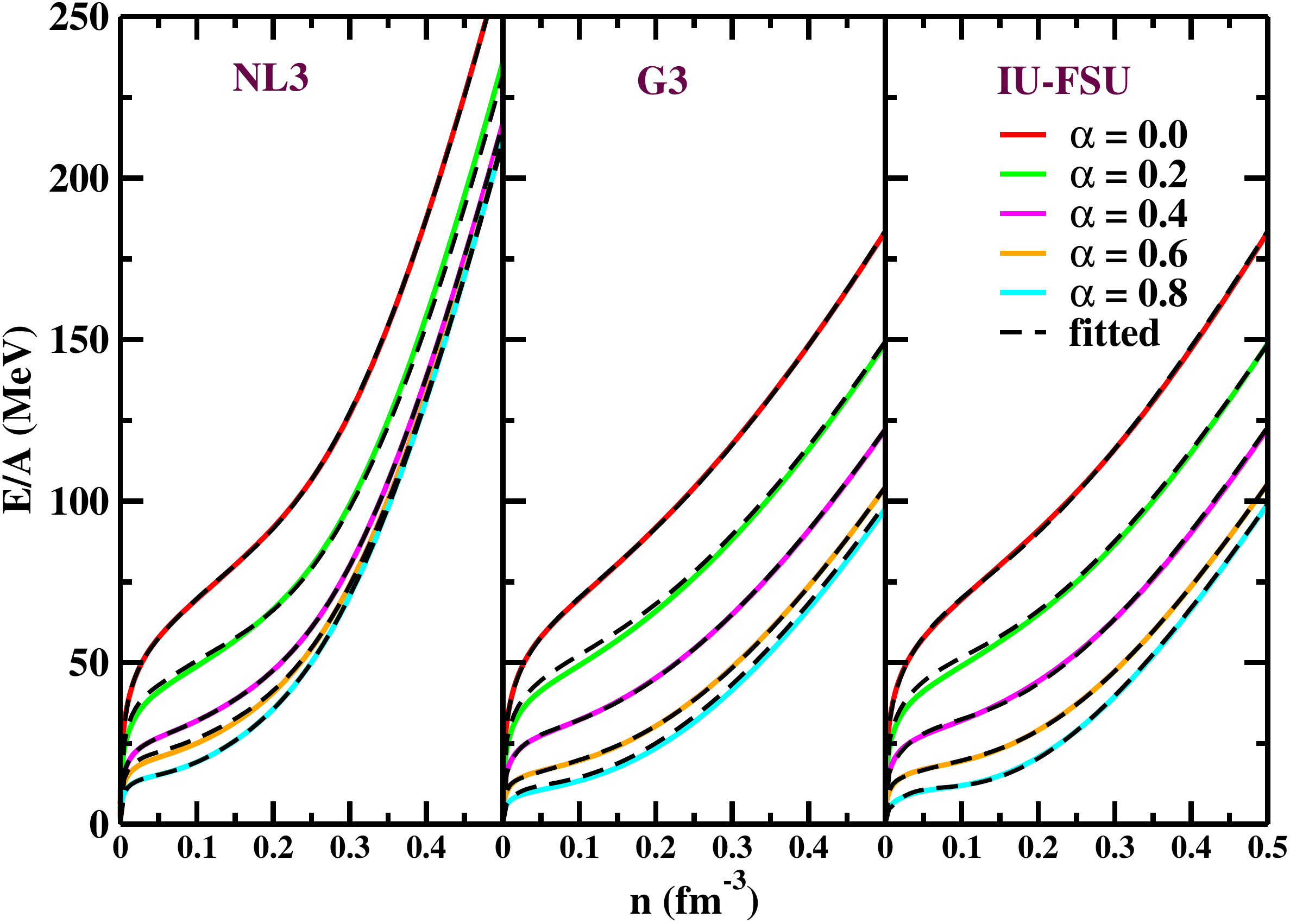}
\caption{(Color online) The neutron star matter saturation curves as a function of baryon number density for different asymmetry $\alpha = \frac{n_n-n_p}{n_n+n_p}$ parameter. The solid curve represents the RMF numerical data and the dotted black curve stands for the fitted expression.}
\label{fitting}
\end{figure}

\begin{figure}
\centering
\includegraphics[width=1\columnwidth]{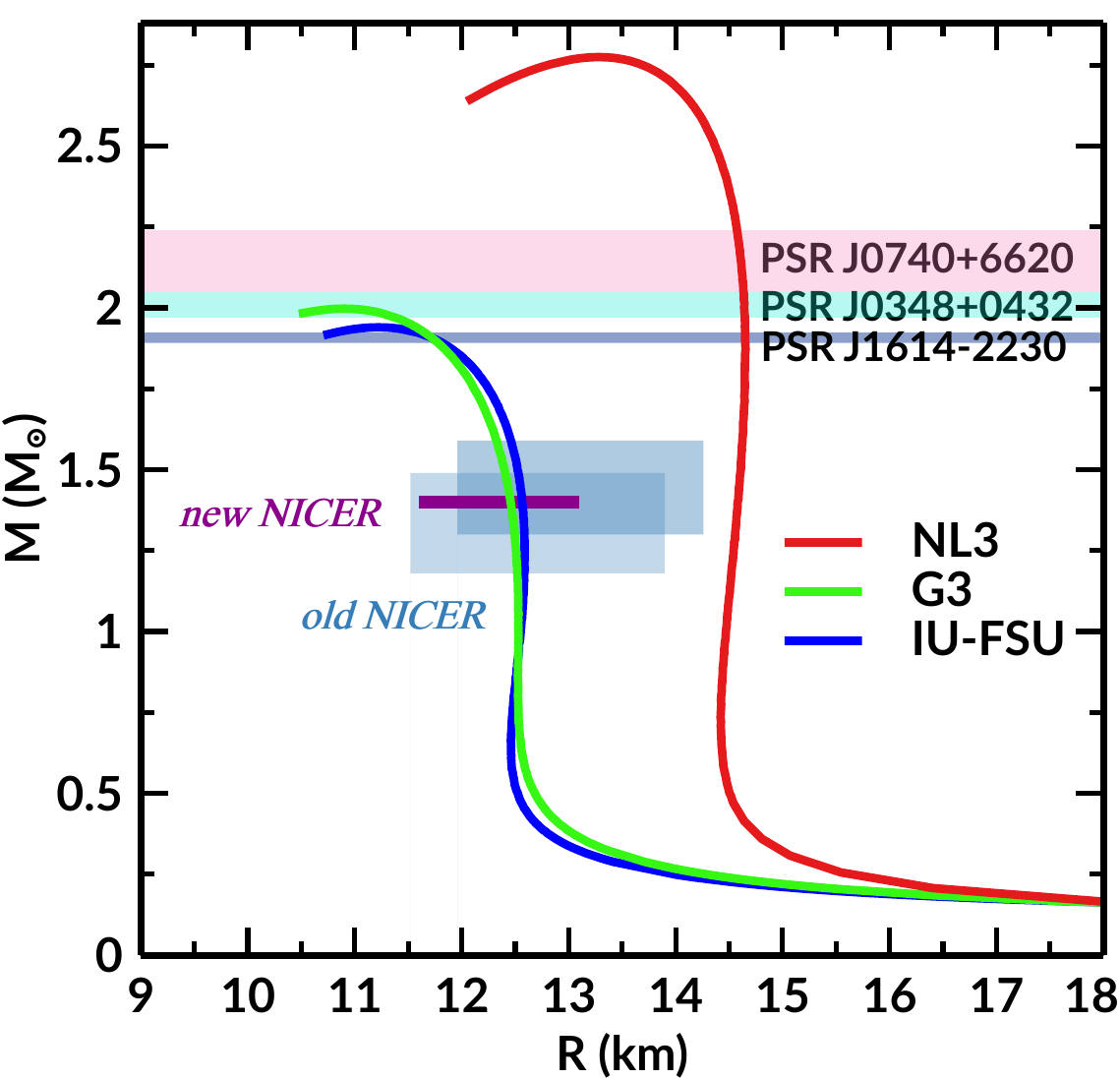}
\vspace{0.1 cm}
\caption{(color online) Mass-radius profile of a neutron star for NL3 (red), G3 (green) and IU-FSU (blue) parameter sets. The old NICER data given in two boxes from the two different analysis \cite{Miller_2019, Riley_2019}. The horizontal line in violet colour represents the new NICER constraint on the radius of the canonical star \cite{miller2021radius}.}
\label{m-rcurve}
\end{figure}

In the last few decades, the theoretical models destined to explore the behaviour of dense nuclear medium has been proved inevitable to unravel the properties of compact astrophysical objects i.e. neutron stars or white dwarfs. The fascinating structural and compositional resemblance of finite nuclei and neutron stars allude us that the physics of compact objects can be explored by extrapolating the data of terrestrial experiments and theoretical formulization of dense matter systems \cite{sym12060898}. From being the extraordinary states of highly dense matter in the inner core to the pasta phases of nuclei at ordinary densities in the outer crust, neutron star manifests the distribution of matter thoroughly \cite{2012ARNPS..62..485L}. It is evidential that the neutron star cores are $10^{3}$ times or more denser than the density at ``neutron drip" line, so, we can utilize a consistent, congruous and ultra-high dense equation of state of nuclear matter for an effectual perception of neutron star properties \cite{PhysRevLett.73.2650}. The comprehensive knowledge of the EoS of nuclear matter and pure neutron matter delineate a prominent bridge between the finite nuclei and dense interstellar bodies. The EoS of strong interacting dense matter is the key component for the determination of general properties of neutron star (maximum mass, radius, tidal deformability) and it also controls the cooling rate and dynamics of core-collapse supernovae remnants \cite{PhysRevC.100.055802, Bombaci_2018}. Immobilizing the correct EoS for the compact stellar objects had been a complex task in the nuclear and astro-particle physics over the last few decades. Several constraints had been enforced on the EoS at high density with the help of observational gravitational wave data (GW170817) \cite{PhysRevLett.119.161101}, Einstein Observatory (HEAO-2) \cite{BOGUTA1981255} and various generations of X-ray radio telescopes \cite{NASA, Greif_2020}. There are many non-relativistic (Skyrme \cite{doi:10.1098/rspa.1961.0018}, Gogny forces \cite{PhysRevC.21.1568}) and relativistic (Relativistic Mean-Field Model \cite{WALECKA1974491, GAMBHIR1990132}) theoretical approaches which endue us with a consistent formalism to construct the EoS and calculate the empirical properties of strongly-interacting dense matter systems, which we can adopt as a manifestation of compact stars. The relativistic class of models are an alternative and more factual approach for low-energy Quantum Chromodynamics with all the built-in non-perturbative properties (i.e. current conservation, local or global symmetry breaking etc.) where baryons and nuclei are stabilized as solitons in a mesonic fluid \cite{ADAM2020135928, Adam_2015}. With the advancement in the cumbersome algebra of quantum field theory, the effective interaction between the nucleons and mesons can be expressed in the form of energy density functionals., which can be approached with the help of self-consistent relativistic mean-field (RMF) model. \\
In this work, we apply the RMF formalism to obtain the EoS for a dense matter system, where along with the neutrons and protons, electrons are also present to maintain the charge neutrality. We will denote this kind of infinite dense matter (consisting neutron, proton and electron) as neutron star matter (NSM) for our further discussion. Since, our aim here is to calculate the properties for a neutron star, so it is a necessity to add electrons and muons in our system to maintain the compositional and neutrality properties of the star. Now, the Lagrangian for such a dense matter within RMF formalism  can be written as \cite{Kumar_2020}
\begin{eqnarray}\label{lag}
{\cal L} & = &  \sum_{i=p,n} \bar\psi_{i}
\Bigg\{\gamma_{\nu}(i\partial^{\nu}-g_{\omega}\omega^{\nu}-\frac{1}{2}g_{\rho}\vec{\tau}_{i}\!\cdot\!\vec{\rho}^{\,\nu})
-(M-g_{\sigma}\sigma\nonumber\\
&&
-g_{\delta}\vec{\tau}_{i}\!\cdot\!\vec{\delta})\Bigg\} \psi_{i}
  +\frac{1}{2}
   \partial^{\nu}\sigma\,\partial_{\nu}\sigma
  -\frac{1}{2}m_{\sigma}^{2}\sigma^2+\frac{\zeta_0}{4!}g_\omega^2
   (\omega^{\nu}\omega_{\nu})^2
   \nonumber \\
& & \null 
-g_{\sigma}\frac{m_{\sigma}^2}{M}
\Bigg(\frac{\kappa_3}{3!}
  + \frac{\kappa_4}{4!}\frac{g_{\sigma}}{M}\sigma\Bigg)
   \sigma^3
+\frac{1}{2}m_{\omega}^{2}\omega^{\nu}\omega_{\nu}
   -\frac{1}{4}F^{\nu\alpha}F_{\nu\alpha}\nonumber\\
 &&  \null
   +\frac{1}{2}\frac{g_{\sigma}\sigma}{M}\Bigg(\eta_1+
 \frac{\eta_2}{2} \frac{g_{\sigma}\sigma}{M}\Bigg)m_\omega^2\omega^{\nu}\omega_{\nu}+\frac{1}{2}\eta_{\rho}\frac{m_{\rho}^2}{M}g_{\sigma}\sigma(\vec\rho^{\,\nu}\!\cdot\!\vec\rho_{\nu}) \nonumber\\
 &&
 +\frac{1}{2}m_{\rho}^{2}\rho^{\nu}\!\cdot\!\rho_{\nu} -\frac{1}{4}\vec R^{\nu\alpha}\!\cdot\!\vec R_{\nu\alpha}
  -\Lambda_{\omega}g_{\omega}^2g_{\rho}^2(\omega^{\nu}\omega_{\nu})(\vec\rho^{\,\nu}\!\cdot\!\vec\rho_{\nu})
 \nonumber\\
 &&
 +\frac{1}{2}\partial^{\nu}\vec\delta\,\partial_{\nu}\vec\delta-\frac{1}{2}m_{\delta}^{2}\vec\delta^{\,2} +  \sum_{l=e,\mu}\bar\phi_{l}\,(i\gamma_{\nu} \partial^{\nu} - m_l)\phi_{l},
\label{eq1}
\end{eqnarray}
where the last term stands for the added electrons and muons in the nuclear matter and $\psi$, $\bar\psi$, $\phi_{l}$ are the wave-functions of the nucleons (proton and neutron), anti-nucleons and leptons respectively. It includes $\sigma$, $\omega$, $\rho$ and $\delta$ mesons to represent the interaction of nucleons, self and cross-coupled interactions. $M$ stands for the mass of the nucleons; $m_{\sigma}$, $m_{\omega}$, $m_{\rho}$, $m_{\delta}$, $g_{\sigma}$, $g_{\omega}$, $g_{\rho}$, $g_{\delta}$ are the masses and the self-coupling constants for $\sigma$, $\omega$, $\rho$ and $\delta$ mesons respectively;  $\kappa_3$, $\kappa_4$, $\zeta_0$, $\eta_1$, $\eta_2$, $\eta_\rho$ and $\Lambda_\omega$ are the coupling constants; $F^{\nu\alpha}$ and $\vec R^{\nu\alpha}$ are field strengths and $\tau_{3}$ is the isospin operator. A more detailed and term by term explanation of the Lagrangian is discussed in the references \cite{Kumar_2020, BharatKumar, KUMAR2017197, Avancini:2002kf}. We obtain the EoS for the defined neutron star matter by applying the Euler-Lagrange's equation of motion and the relativistic mean field approximation on the above Lagrangian ( Eq. \ref{eq1}) \cite{BharatKumar, PhysRevC.89.044001}. To get the comprehensive idea of the EoS, we derive the EoS for the complete range of asymmetry factor. The asymmetry parameter ($\alpha$) is defined as $\alpha = \frac{n_n-n_p}{n_n+n_p}$, with $n_{n}$ and $n_{p}$ being the neutron and proton density, and $\alpha = 1$ stands for purely neutron matter. The number density of electrons and muons in the neutron star matter is kept equal to the proton number density to maintain the charge neutrality i.e.,\\
\begin{eqnarray} \label{eqm}
n_{p} &=& n_{e} + n_{\mu}.
\end{eqnarray}
The binding energy per nucleon ($E/n - M$, $E$ being the energy density and $n$ is the total nucleon density) curve as a function of nucleonic number density is depicted in Fig. \ref{fitting}.\\
In the present work, we used three different RMF parameter sets (NL3\cite{PhysRevC.55.540}, G3\cite{KUMAR2017197} and IU-FSU\cite{PhysRevC.100.025805}), with NL3 being the stiffest and the recently developed G3 being the soft, compose the whole range of equation of state. The values of nuclear matter properties (i.e. saturation density, binding energy, incompressibility etc.) for symmetric nuclear matter predicted by the chosen RMF parameter sets satisfy all the empirical/experimental constraints. The numerical magnitude of incompressibility of symmetric nuclear matter ($K$) at saturation density for G3 and IU-FSU forces are $244$ and $231$ $MeV$, and that of symmetry energy ($S$) are $31$ and $32$ $MeV$ respectively \cite{Kumar_2020, HCDas}, which are appropriate for the range stated by various theoretical and experimental models \cite{GARG201855, DANIELEWICZ20141}. The three considered parameter sets are widely used in literature and our aim is to show the variation of the results with different forces. It is to be noted here that NL3 is one of the most successful parameter set for finite nuclei. This set is also used to explain the GW190814 data with an admixture of dark matter inside the neutron star \cite{DasPRD_2021}. The coupling constants and the nuclear matter properties at saturation density along with the empirical/experimental values are presented in the table \ref{table1}. \\
\begin{table} 
\label{table1}
\caption{The coupling constants and the nuclear matter properties at saturation for the EoS of NL3 \cite{PhysRevC.55.540}, G3 \cite{KUMAR2017197} and IU-FSU \cite{PhysRevC.100.025805} parameter sets. The nucleon mass ($M$) is 939.0 MeV.  All of the coupling parameters are dimensionless and the NM parameters are in MeV, except $n_{0}$ which is in fm$^{-3}$. The NM parameters are given at saturation density for NL3, G3 and IU-FSU parameter sets in the lower panel. The references are $[a]$,$[b]$, $[c]$ $\&$ $[d]$ \cite{Zyla_2020}, $[e] $\&$ [f]$ \cite{doi:10.1146/annurev.ns.21.120171.000521}, $[g]$ \cite{GARG201855}, $[h] $\&$ [i]$ \cite{DANIELEWICZ20141}, and $[j]$ \cite{zimmerman2020measuring}.}
\begin{tabular}{cccccccccc}
\hline
\hline
\multicolumn{1}{c}{Parameter}
&\multicolumn{1}{c}{NL3}
&\multicolumn{1}{c}{G3}
&\multicolumn{1}{c}{IU-FSU}
&\multicolumn{1}{c}{Empirical/Expt. Value}\\
\hline
$m_{\sigma}/M$ & 0.541 & 0.559 & 0.523 &0.426 -- 0.745 $[a]$\\
$m_{\omega}/M$  &  0.833  &  0.832 & 0.833 & 0.833 -- 0.834 $[b]$  \\
$m_{\rho}/M$  &  0.812  &  0.820 & 0.812 & 0.825 -- 0.826 $[c]$\\
$m_{\delta}/M$   & 0.0  &   1.043 & 0.0 & 1.022 -- 1.064 $[d]$\\
$g_{\sigma}/4 \pi$  &  0.813  &  0.782 & 0.793 & \\
$g_{\omega}/4 \pi$  &  1.024  &  0.923 & 1.037 &\\
$g_{\rho}/4 \pi$  &  0.712 &  0.962  & 1.081 & \\
$g_{\delta}/4 \pi$  &  0.0  &  0.160 & 0.0 & \\
$k_{3} $   &  1.465 &    2.606 & 1.1593 &  \\
$k_{4}$  &  -5.688  & 1.694   & 0.0966 &\\
$\zeta_{0}$  &  0.0 &  1.010 & 0.03 &   \\
$\eta_{1}$  &  0.0 &  0.424 & 0.0 & \\
$\eta_{2}$  &  0.0 &  0.114 & 0.0 &  \\
$\eta_{\rho}$  &  0.0 &  0.645 & 0.0 & \\
$\Lambda_{\omega}$  &  0.0 &  0.038 & 0.046 &  \\
\hline
$n_{0}$ & 0.148 & 0.148 & 0.154 & 0.148 -- 0.185 $[e]$\\
$B.E.$ & -16.29 & -16.02 & -16.39 & -15.00 -- 17.00 $[f]$\\
$K$ & 271.38 & 243.96 & 231.31 &220 -- 260 $[g]$\\
$S$ & 37.43 & 31.84 & 32.71 & 30.20 -- 33.70 $[h]$\\
$L_{sym}$ & 120.65 & 49.31 & 49.26 & 35.00 -- 70.00 $[i]$ \\
$K_{sym}$ & 101.34 & -106.07 & 23.28 & -174 -- -31 $[j]$\\
$Q_{sym}$ & 177.90 & 915.47 & 536.46 & -----------\\
\hline
\hline
\end{tabular}
\end{table}
To achieve an expressional form of energy functional for the effective interactions in the neutron star matter explained by RMF formalism, we have fitted the numerically obtained data. The assumptive form of the fitted energy functional is motivated by the work of  Br$\ddot{u}$ckner {\it et al.} \cite{PhysRev.168.1184, PhysRev.171.1188}. There were several issues with the  Br$\ddot{u}$ckner's energy functional, for instance, it can not rectify the Coester-band problem \cite{PhysRevC.5.1135} and it is defined purely for the non-relativistic nuclear matter formalism. We ameliorate the  Br$\ddot{u}$ckner's energy functional in the local density approximation, so that, it can satisfy the RMF data \cite{PhysRevC.103.024305}, and also added some series of potential function and an extra term for the lepton's kinetic energy inclusion. The modified energy density functional can be stated as,
\begin{eqnarray} \label{efitting}
{\cal E} & = & C_k n^{2/3} + C_e n^{4/9} + \sum_{i=3}^{14} (b_i + a_i \alpha^2) n^{i/3},
\end{eqnarray}
where $C_{k} = 0.3 (\hbar^{2}/2M) (3\pi^{2})^{2/3} [(1+\alpha)^{5/3} + (1-\alpha)^{5/3}]$ \cite{PhysRev.171.1188} is the kinetic energy coefficient for nucleons and $C_{e} = b_{e} (1-\alpha)^{5/9}$, with $b_{e}$ as a variable obtained during fitting procedure, is the kinetic term coefficient for leptons. The last term stands for the potential interaction of the nucleons and the coefficients $b_{i}$ and $a_{i}$ has to be obtained by fitting procedure for different RMF parameter sets. We observed that the accuracy of the fitting mechanism decreases if we reduce the number of coefficients ($a_{i}$ and $b_{i}$) in the expansion of potential term of Eq. \ref{efitting} \cite{PhysRevC.103.024305}. The mean deviation `$\delta$', which is defined as, $\delta=[\sum_{j=1}^{N} (E/A)_{j,\mathrm{Fitted}} - (E/A)_{j,\mathrm{RMF}}]/N$, N being the total number of points, is $18\%$, $6\%$ and $0.5\%$ for 8 (i.e. `$i$' runs from $3$ to $10$ in Eq. \ref{efitting}), 10 and 12 terms respectively. The fitted curves of the neutron star matter with the above expression (Eq. \ref{efitting}) are shown in the Fig. \ref{fitting} through the black dotted lines. We have used this energy density functional as an input in the coherent density fluctuation method to achieve the range of certain properties of neutron star (incompressibility, symmetry energy and slope parameter) for the first time. Coherent density fluctuation model (CDFM) was first introduced three decades ago by Antonov {\it et al.} and is now a well established formalism to decipher the properties of finite nuclei \cite{Antonov1980, PhysRevC.50.164, Gaidarov2020ProtonAN}. The mere backbone of the CDFM model is that a coordinate generator  `$x$' can be used to write the one-body density matrix $n(r,r^\prime)$ of a nucleus as a sequential superposition of infinite number of one-body density matrices $n_{x}(r,r^\prime)$, which are coined as ``fluctons" \cite{PhysRevC.84.034316, Kaur2020OnTS}. The density of fluctons has the form,
\begin{equation}
n_x ({\bf r}) = n_0 (x)\, \Theta (x - \vert {\bf r} \vert),
\label{denx}
\end{equation}
where $n_{0}(x)$ is defined as $n_{0}(x) = 3 A / 4 \pi x^{3}$, $A$, being the total number of nucleons in the finite matter. Within the CDFM approach, the density distribution  of the spherical finite nuclear matter of radius `$r$' can be expressed as \cite{Antonov_2018, Antonov_2016},
\begin{eqnarray}
n (r) &=& \int_0^{\infty} dx\, \vert F(x) \vert^2\, n_{0}(x) \, \Theta(x-\vert{\bf r} \vert),
\label{rhor}
\end{eqnarray}
$\vert F(x) \vert^2$ is defined as the weight function and it can be obtained theoretically for a monotonically decreasing local density in the generator coordinate `$x$' as \cite{Antonov_2018},
\begin{equation}
|F(x)|^2 = - \frac{1}{n_0 (x)} \frac{dn (r)}{dr} \Bigg|_{r=x},
\label{wfn}
\end{equation}
Now, the nuclear and structural properties of the neutron star, i.e. incompressibility ($K^{star}$), symmetry energy ($S^{star}$), slope parameter ($L^{star}_{sym}$) and curvature ($K^{star}_{sym}$), can be expressed in terms of weight function and the expression of the parameters evaluated from the energy density functional (eq. \ref{efitting}) of infinite star matter system, as, \cite{PhysRevC.84.034316, PhysRevC.85.064319, Antonov_2016}
\begin{eqnarray}
K^{star} &=& \int_0^{\infty} dx\, \vert F(x) \vert^2 \ K^{NSM} (n (x)),
\label{K0} \\
S^{star} &=& \int_0^{\infty} dx\, \vert F(x) \vert^2\, S^{NSM} (n (x)) ,
\label{s0} \\
L_{sym}^{star} &=& \int_0^{\infty} dx\, \vert F(x) \vert^2 \,L_{sym}^{NSM} (n (x)) ,
\label{L0} \\
K_{sym}^{star} &=& \int_0^{\infty} dx\, \vert F(x) \vert^2 \ K_{sym}^{NSM} (n (x)),
\label{k0}
\end{eqnarray}
The profile of the weight function with density holds the key information about the dependence of the calculated properties on the structure and composition of the defined matter system. The magnitude of the weight function at a value of `$x$' will decide the share of that particular region of density in the overall magnitude of calculated property. The energy density functional for the neutron star matter can be converted from momentum space to the coordinate space `$x$' in a local density approximation technique using the Br$\ddot{u}$ckner method.The expressions for $K^{NSM}$, $S^{NSM}$, $L^{NSM}_{sym}$ and $K^{NSM}_{sym}$ can be obtained from Eq. \ref{efitting}, by applying their common derivative definitions \cite{Fetter, PhysRevC.80.014322, PhysRevC.90.044305}, i.e.,
the NM parameters $K^{NM}$, $S^{NM}$, $L_{sym}^{NM}$ and $K_{sym}^{NM}$ are obtained from the following standard relations:
\begin{eqnarray}
K^{NM}&=&9\rho_0^2\frac{\partial^2}{\partial \rho^2} \bigg(\frac{\cal E}{\rho}\bigg)\Big|_{\rho=\rho_0} \label{knm},\\
S^{NM}&=&\frac{1}{2}\frac{\partial^2 ({\cal E}/\rho)}{\partial\alpha^2}\Big|_{\alpha=0},\label{snm}\\
L_{sym}^{NM}&=&3\rho_0\frac{\partial S(\rho)}{\partial\rho}\Big|_{\rho=\rho_0} = \frac{3P}{\rho_{0}},\label{lsymnm}\\
K_{sym}^{NM}&=&9\rho_0^2\frac{\partial^2 S(\rho)}{\partial\rho^2}\Big|_{\rho=\rho_0}.\label{ksymnm}
\end{eqnarray}
which are given as follow using Eq. (3)
\begin{eqnarray}
K^{NSM} &=& -150.12\,n_0^{2/3}(x) - 2.22\,b_{e}\,n_{0}^{4/9}(x) \nonumber \\ 
&&
+ \sum_{i=4}^{14} i\, (i-3)\, b_i\, n_0^{i/3}(x), \label{eqA}\\
S^{NSM} &=& 41.7\,n_0^{2/3}(x) - 0.12\,b_{e}\,n_{0}^{4/9}(x) \nonumber \\
&&
+ \sum_{i=3}^{14} a_i\, n_0^{i/3}(x), \label{eqB}\\
L_{sym}^{NSM} &=& 83.4\,n_0^{2/3}(x) - 0.16\,b_{e}\,n_{0}^{4/9}(x) \nonumber \\
&&
+ \sum_{i=3}^{14} i\, a_i\, n_0^{i/3}(x), \label{eqC}\\
K_{sym}^{NSM} &=& -83.4\,n_0^{2/3}(x) + 0.266\,b_{e}\,n_{0}^{4/9}(x) \nonumber \\
&&
+ \sum_{i=4}^{14} i\, (i-3)\, a_i\, n_0^{i/3}(x), \label{eqD}
\end{eqnarray}
The above expressions are derived for symmetric ($\alpha = 0$) case. Now, we can easily calculate the nuclear properties of the neutron star with the help of weight function and using the above expressions. The weight function computation demand the slope of density curve with respect to radius of the neutron star. We compute the mass-radius profile and the density curve of the neutron star for the three considered RMF parameter sets. The density curve and the mass-radius profile of the neutron star can be acquired by imposing the beta equilibrium conditions \cite{1965MmSAI36323P, Kumar_2020} in the neutron star matter and  using the obtained EoS as an input for the Tolman-Oppenheimer-Volkoff (TOV) equations \cite{PhysRev.55.374, PhysRev.55.364}.  The $\beta$- equilibrium conditions and the TOV equations for the static isotropic proto-neutron star can be written as, \\
\begin{eqnarray} \label{TOV}
\mu_{n} &=& \mu_{p} + \mu_{e}, \nonumber \\
\mu_{e} &=& \mu_{\mu}, \\ \nonumber \\
\mathrm{and} \nonumber \\  \nonumber \\
\frac{d P(r)}{d r} &=&  -\frac{[E(r)+P(r)]}{r^2\Big(1-\frac{2M(r)}{ r}\Big)} [M(r)+{4\pi r^3 P(r)}], \nonumber \\
\frac{d M(r)}{d r} &=& 4\pi r^2 {E(r)}.
\end{eqnarray}
Here $\mu_{n}$, $\mu_{p}$, $\mu_{e}$, and $\mu_{\mu}$ describe the chemical potentials of the neutrons, protons, electrons, and muons respectively; $E$ and $P$ are the energy density and pressure of the neutron star. The self consistent numerical solution of Eq. (2) and Eq. (15) will  set  the  fraction  of  neutron,  proton,  electron  and  muon number density for a given baryon density in a neutron star. $M(r)$ is defined as the mass of the neutron star at radius r and the boundary conditions to solve these equations are $P(R) = 0$, for a particular choice of central density $n_c = n(0)$. Finally, we also added the crust part in the above computed EoS to get a detailed and complete analysis of the neutron star properties. We extended the surface part of the NS mathematically by adding the crust energy and pressure calculated by Baym, Pethick and Sutherland, i.e. BPS crust EoS, in the tail part of all the three RMF parameter's main equation of state \cite{1971ApJ...170..299B}. A more detailed formalism to calculate the mass-radius profile of the neutron star using RMF equation of state can be found in the references \cite{Kumar_2020, BharatKumar, 10.1093/mnras/staa1435}. The mass-radius profile for all the three assumed parameter sets is shown in fig. \ref{m-rcurve}.\\
The maximum mass of the neutron star calculated with the help of G3 and IU-FSU parameter sets are $2.004 M_\odot$ and $1.940 M_\odot$ respectively \cite{Kumar_2020, Das_2021}, which fit well in the range of the observational pulsar data PSR J1614-2230 ($M=1.908\pm0.04M_\odot$) \cite{Arzoumanian_2018}, PSR J0348+0432 ($M=2.01\pm0.04 \ M_\odot)$ \cite{Antoniadis_2013} and PSR J0740+6620 ($M=2.15^{+0.10}_{-0.09}M_\odot$) \cite{Cromartie_2019}. Recent evaluation of the pulsar data PSR J0740+6620 done by Fonseca et. al. enumerate the mass of the star in the range $2.08\pm0.07M_\odot$ with $68\%$ confidence limit \cite{Fonseca_2021}. The old NICER \cite{Miller_2019, Riley_2019} data are satisfied by both G3 and IU-FSU sets. Also, recently the new equatorial circumferential radius measurements are reported by Miller at. al. \cite{miller2021radius} on the basis of NICER and XMM-Newton X-ray observation of PSR J0740+6620 within $68\%$ confidence limit. However, the mass predicted by the NL3 parameter set is quite larger than the defined limit. The radius constraint put recently by the Miller et. al. using NICER simulations for the canonical star ($1.4 M_\odot$) is also well satisfied by the G3 and IU-FSU parameter sets \cite{miller2021radius}. Also, the constraint set by the observational data of GW170817 event for the tidal deformability of canonical star is satisfied by the G3 parameter set ($\Lambda = 582.26 $) \cite{PhysRevC.97.045806, PhysRevLett.119.161101, das2021effects}. So, we can claim that the assumed RMF parameter sets are consistent with the astrophysical observational data and well-suited to calculate the various properties of neutron star.\\   
\begin{figure}
\centering
\includegraphics[width=1\columnwidth]{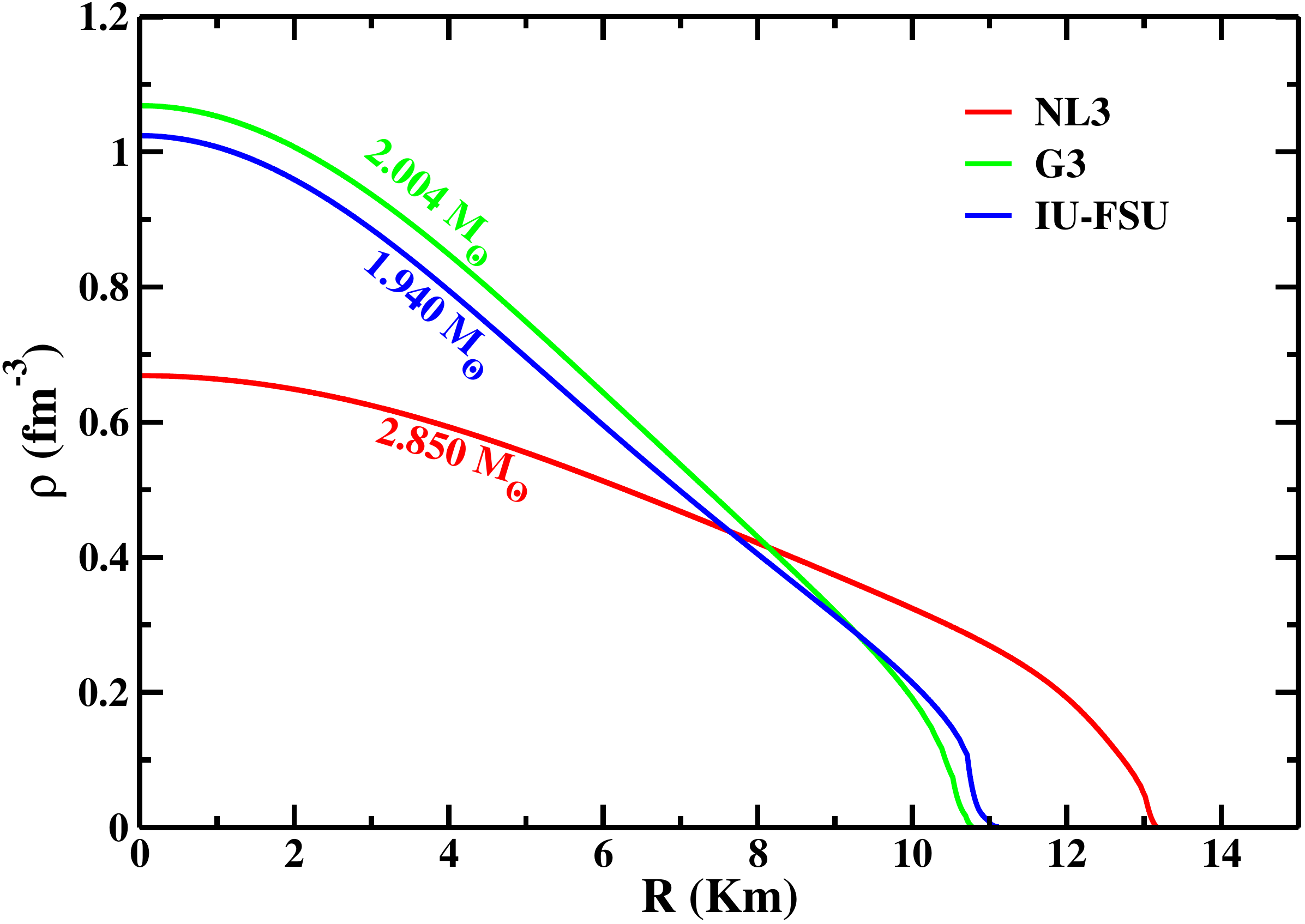}
\vspace{0.1 cm}
\caption{(color online) The neutron star densities ($\rho$) for NL3 (red), G3 (green) and IU-FSU (blue) parameter sets as a function of radius of the maximum mass star (R). The mass number ($A$) of the maximum mass neutron star for NL3, G3 and IU-FSU are $3.35\times10^{54}$, $2.32\times10^{54}$ and $2.23\times10^{54}$ respectively.  }
\label{d-rcurve}
\end{figure}
The density-radius curve of the neutron star for NL3, G3 and IU-FSU parameter sets is depicted in Fig. \ref{d-rcurve}. The density-radius curve is computed for the maximum mass predicted by the corresponding parameter set i.e. $2.850 M_\odot$ for NL3, $2.004 M_\odot$ for G3 and $1.940 M_\odot$ for IU-FSU. We observe that the central density of the neutron star is maximum for the G3 parameter set, while NL3 being the stiffest EoS, have the lowest central density.\\
However, with the help of the density-radius curve, we can calculate the weight function ($\vert F(x) \vert^2$) of the neutron star. The total number of nucleons for the neutron star with the maximum mass predicted by NL3, G3 and IU-FSU parameter sets with mass number A are $3.35\times10^{54}$, $2.32\times10^{54}$ and $2.23\times10^{54}$ respectively \cite{Glendenning:1997wn}. The total number of nucleons computed for a canonical star are $1.53297\times10^{54}$, 1.53281$\times 10^{54}$, and 1.53286 $\times 10^{54}$ respectively for NL3, G3 and IU-FSU sets. Although, the number difference appears after the third decimal for all three forces, actually these numbers are quite different from each other due to the order of magnitude. It is worth mentioning that the neutron star is a big nucleus with nucleons, electrons and muons, which has variation of density with radius as shown in Fig. 3. It possesses all properties of a finite nucleus with mass number A. The giant monopole excitation energy controls by the incompressibility of the nucleus \cite{PhysRevC.65.054323, PATRA200167, PATRA2002240}. Here also the $K^{star}$ gives significant information about the giant resonances of the neutron star.  
\begin{table*}
\label{table2}
\caption{The numerical values of incompressibility, symmetric  energy, slope  and curvature parameter for maximum mass and the canonical mass star of the corresponding RMF parameter sets. The maximum mass for NL3, G3 and IU-FSU parameter sets are $2.850 M_\odot$, $2.004 M_\odot$ and $1.940 M_\odot$ respectively. All the values are in MeV unit.}
\renewcommand{\tabcolsep}{0.28cm}
\renewcommand{\arraystretch}{2.0}
\label{table2}
\begin{tabular}{|c|c|c|c|c|c|c|}
\hline
\multirow{2}{*}{Parameter Set} & \multicolumn{2}{c|}{NL3} & \multicolumn{2}{c|}{G3} & \multicolumn{2}{c|}{IU-FSU} \\ \cline{2-7} 
 & \begin{tabular}[c]{@{}c@{}}Maximum mass\end{tabular} & \begin{tabular}[c]{@{}c@{}}Canonical  mass\end{tabular} & \begin{tabular}[c]{@{}c@{}}Maximum mass\end{tabular} & \begin{tabular}[c]{@{}c@{}}Canonical mass\end{tabular} & \begin{tabular}[c]{@{}c@{}}Maximum mass\end{tabular} & \begin{tabular}[c]{@{}c@{}}Canonical mass\end{tabular} \\ \hline
$K^{star}$ & 44.956 & 3.934 & 29.480 & 3.590 & 29.827 & 3.389 \\ \hline
$S^{star}$ & 146.002 & 15.911 & 66.813 & 8.904 & 60.758 & 7.228 \\ \hline
$L^{star}_{sym}$ & 615.854 & 65.038 & 307.015 & 45.392 & 320.321 & 48.225 \\ \hline
$K^{star}_{sym}$ & -688.514 & -70.534 & -360.127 & -44.315 & -228.012 & -26.282 \\ \hline
\end{tabular}
\end{table*}
With all the input ingredients acquired, we calculate the numerical values of incompressibility, symmetry energy and its derivatives for maximum mass neutron star and canonical star represented by all the three parameter sets using RMF density functional and CDFM model. Till now, to the best of our knowledge we did not find any work in the literature regarding the availability of exact numerical values of nuclear matter properties of neutron star or any theoretical model which can endue us with such formalism. We, here, are explicating the numerical values of incompressibility, symmetry energy, slope and curvature parameter (Table \ref{table2}) of a neutron star for the chosen RMF forces, which is unaccustomed. We deciphered some interesting results from the values of Table \ref{table2}. We observed that the values of all the nuclear properties for maximum mass star with NL3 is greater in comparison to G3 and IU-FSU forces. The magnitude of incompressibility coefficient for maximum mass star with NL3, G3 and IU-FSU parameter sets are $44.956$, $29.480$ and $29.827$ $MeV$ respectively. This tendency of NL3 predicting the higher values of incompressibility and other properties for neutron star justify its nature of stiff EoS, which has also been anticipated for symmetric nuclear matter case \cite{PhysRevC.55.540}.

Another important dimension of incompressibility is that the incompressibility of the matter decreases with increase of asymmetricity and density. For example, the incompressibility of pure nuclear matter, i.e., with equal number of protons and neutrons at saturation is 243.96 MeV for G3 set and it is 29.827 MeV for neutron star matter, which has a large asymmetricity and density. This phenomenon of decrement in the incompressibility coefficient as we keep increasing the density of the system has also been noticed and reported in the literature priorly \cite{Kumar_2020}. The validity of a equation of state can be solely checked by computing its incompressibility coefficient. The numerical range for incompressibility coefficient is indeed the most important quantity to calculate as it restricts the stiffness of the equation of state of the system by checking the compatibility of the equation of state with causality, which require the adiabatic sound speed not to exceed the speed of light \cite{10.1093/mnras/staa1435}. The lowest central density for NL3 parameter set despite being the prediction of highest mass is the result of causality restriction, as the stiffness of the equation of state which is related to the incompressibility should be be compatible with causality at highest density \cite{PhysRevC.63.015802}. We realise that the maximally incompressible equation of state can be soften by reducing the $K$, which in turn will reduce the maximum neutron star mass significantly. To inspect the shift of incompressibility coefficient with the mass of the neutron star, we extend the calculations for different masses using G3 parameter set. We observe that the values of $K^{star}$ with G3, as a representative parameter set, for $1.4M_\odot$, $1.6M_\odot$ , $1.7M_\odot$, $1.8M_\odot$ and $2.004M_\odot$ are $3.590$, $5.203$, $6.302$, $9.163$ and $29.480$ MeV respectively. However, on the other hand, the incompressibility coefficient of the canonical star is almost same for all the considered RMF parameter sets, given in table \ref{table2}. This particular observation of the incompressibility coefficient seems to indicate that the value of $K^{star}$ is proportional to the mass of the star, which sequentially depends on the number of nucleons. A more detailed study regarding the observance and conclusions of the correlations between the incompressibility and mass of the neutron star is in progress and will be published somewhere else \cite{Pattnaik}.

The magnitude of the symmetry energy for maximum mass neutron star is also a bit higher in comparison to the magnitude of symmetric nuclear matter at saturation density for all the RMF parameter sets. The numerical values of symmetry energy at saturation density for a symmetric nuclear matter with NL3, G3 and IU-FSU forces are $37.43$, $31.84$ and $32.71$ MeV respectively, while those for the case of maximum mass neutron star comes out to be $146.002$, $66.813$ and $60.758$ MeV. The values of symmetry energy for canonical star is quite smaller in magnitude as compare to maximum mass star of the corresponding RMF parameter set. Also, contrary to the case of incompressibility, in spite of being the same mass of canonical star for all the three parameter sets, the magnitude of the symmetry energy is not equal for different parameter set. This kind of behaviour reflects the dependence of symmetry energy on the structure and composition of the neutron star. The incompressibility $K$ is obtained from the derivative of energy density with respect to density, but the symmetry energy is derived from the derivative of the energy density with respect to asymmetricity. Thus, shows a significant variation in the symmetry energy as compared to incompressibility. As we can see from fig. \ref{m-rcurve}, the radius of canonical star differ for the NL3, G3 and IU-FSU parameters, which cause the change in $S^{star}$ for the same mass of the star. 

Although there is no empirical or experimental data available to support the magnitude of the neutron star's symmetry energy, but as we know that neutron star is a highly asymmetric dense object, so, a major change in the symmetry energy with mass is expected due to its isospin-dependent characteristics and the structural behaviour. The unexpectedly larger value of $L_{sym}^{star}$ for NL3 parameter set is also supported by the inclination of stiff EoS of the dense matter towards higher value of slope parameter \cite{sym12060898}. A precise knowledge of the range of symmetry energy and $L_{sym}$ is enough to estimate the radius of a neutron star quite perfectly. A well defined information about the range of symmetry energy and $L_{sym}$ is relevant to trace a more strong and constrained correlation between the surface and volume symmetry energy terms in the mass formula \cite{2015AIPC.1645...61L}. The static dipole polarizability. quadrupole polarizability and the neutron skin thickness is closely related to the correlation between symmetry energy and slope parameter \cite{2015AIPC.1645...61L,Lattimer_2013}. Similar to the case of $S^{star}$, the magnitude of slope parameter for canonical star is also significantly smaller in comparison to the maximum mass star for the considered RMF parameter sets. $K_{sym}^{star}$, being the second order derivative of symmetry energy, is most sensitive and ambiguous quantity to calculate precisely. The negative magnitude of the curvature parameter corroborated by the 1-$\sigma$ constraint and $90\%$ confident bounds on its value at saturation density of symmetric nuclear matter, derived by Josef Zimmerman et al. using the observational data of PSR J0030+0451 and GW170817 event \cite{zimmerman2020measuring, Riley_2019, PhysRevLett.119.161101}. Although, these bounds are not well suited to discuss the curvature parameter of a neutron star, but it however, implies the possibility of negative magnitude and allude the proximity range around the $90\%$ confidence limit of astrophysical observation data. The separation of the contribution of isovector incompressibility or curvature parameter ($K_{sym}$) part from the total incompressibility of a matter can be proved very useful for some teresterrial experiments related to exotic nuclei and heavy-ion collisions \cite{PhysRevC.76.054316, PhysRevLett.102.122502}. A more confident theoretical bound on the value of $K_{sym}$ estimated by the present study through RMF parameter sets for neutron star will path a better way to tune the experimental techniques related to isoscalar giant resonances towards the prediction of properties related to astrophysical objects. A more detailed study about the importance and correlations of the newly introduced parameters for the neutron star ($X^{star}$) are quite interesting and will be published in a future work \cite{Pattnaik}. \\
Despite the deprivation of direct experimental measurement or empirically acquirable data for the nuclear properties i.e. incompressibility, symmetry energy etc. of the neutron star, the numerical values calculated here with the help of consolidated RMF and CDFM formalism appear justifiable and irreproachable. The competency of the present theoretical perspective can be exuberantly validated using various consistent energy density functionals and relevant RMF parameter sets. The accomplished accessibility to neutron star properties through a finite nuclei approach favour more dimensions of strongly correlated bridge betwixt the two unequal size objects. The theoretical approach employed in this work present new opportunities for the nuclear and astrophysicist to unearth the wealth of information on the dense astronautical objects and exotic finite nuclei.       
\bibliography{cdfm.bib}

\begin{thebibliography}{86}%
\makeatletter
\providecommand \@ifxundefined [1]{%
 \@ifx{#1\undefined}
}%
\providecommand \@ifnum [1]{%
 \ifnum #1\expandafter \@firstoftwo
 \else \expandafter \@secondoftwo
 \fi
}%
\providecommand \@ifx [1]{%
 \ifx #1\expandafter \@firstoftwo
 \else \expandafter \@secondoftwo
 \fi
}%
\providecommand \natexlab [1]{#1}%
\providecommand \enquote  [1]{``#1''}%
\providecommand \bibnamefont  [1]{#1}%
\providecommand \bibfnamefont [1]{#1}%
\providecommand \citenamefont [1]{#1}%
\providecommand \href@noop [0]{\@secondoftwo}%
\providecommand \href [0]{\begingroup \@sanitize@url \@href}%
\providecommand \@href[1]{\@@startlink{#1}\@@href}%
\providecommand \@@href[1]{\endgroup#1\@@endlink}%
\providecommand \@sanitize@url [0]{\catcode `\\12\catcode `\$12\catcode
  `\&12\catcode `\#12\catcode `\^12\catcode `\_12\catcode `\%12\relax}%
\providecommand \@@startlink[1]{}%
\providecommand \@@endlink[0]{}%
\providecommand \url  [0]{\begingroup\@sanitize@url \@url }%
\providecommand \@url [1]{\endgroup\@href {#1}{\urlprefix }}%
\providecommand \urlprefix  [0]{URL }%
\providecommand \Eprint [0]{\href }%
\providecommand \doibase [0]{http://dx.doi.org/}%
\providecommand \selectlanguage [0]{\@gobble}%
\providecommand \bibinfo  [0]{\@secondoftwo}%
\providecommand \bibfield  [0]{\@secondoftwo}%
\providecommand \translation [1]{[#1]}%
\providecommand \BibitemOpen [0]{}%
\providecommand \bibitemStop [0]{}%
\providecommand \bibitemNoStop [0]{.\EOS\space}%
\providecommand \EOS [0]{\spacefactor3000\relax}%
\providecommand \BibitemShut  [1]{\csname bibitem#1\endcsname}%
\let\auto@bib@innerbib\@empty
\bibitem [{\citenamefont {Lattimer}\ and\ \citenamefont
  {Prakash}(2004)}]{Lattimer536}%
  \BibitemOpen
  \bibfield  {author} {\bibinfo {author} {\bibfnamefont {J.~M.}\ \bibnamefont
  {Lattimer}}\ and\ \bibinfo {author} {\bibfnamefont {M.}~\bibnamefont
  {Prakash}},\ }\href {\doibase 10.1126/science.1090720} {\bibfield  {journal}
  {\bibinfo  {journal} {Science}\ }\textbf {\bibinfo {volume} {304}},\ \bibinfo
  {pages} {536} (\bibinfo {year} {2004})}\BibitemShut {NoStop}%
\bibitem [{\citenamefont {Lattimer}(2014)}]{LATTIMER2014276}%
  \BibitemOpen
  \bibfield  {author} {\bibinfo {author} {\bibfnamefont {J.~M.}\ \bibnamefont
  {Lattimer}},\ }\href {\doibase
  https://doi.org/10.1016/j.nuclphysa.2014.04.008} {\bibfield  {journal}
  {\bibinfo  {journal} {Nuclear Physics A}\ }\textbf {\bibinfo {volume}
  {928}},\ \bibinfo {pages} {276} (\bibinfo {year} {2014})},\ \bibinfo {note}
  {special Issue Dedicated to the Memory of Gerald E Brown
  (1926-2013)}\BibitemShut {NoStop}%
\bibitem [{\citenamefont {Alam}\ \emph {et~al.}(2016)\citenamefont {Alam},
  \citenamefont {Agrawal}, \citenamefont {Fortin}, \citenamefont {Pais},
  \citenamefont {Provid\^encia}, \citenamefont {Raduta},\ and\ \citenamefont
  {Sulaksono}}]{PhysRevC.94.052801}%
  \BibitemOpen
  \bibfield  {author} {\bibinfo {author} {\bibfnamefont {N.}~\bibnamefont
  {Alam}}, \bibinfo {author} {\bibfnamefont {B.~K.}\ \bibnamefont {Agrawal}},
  \bibinfo {author} {\bibfnamefont {M.}~\bibnamefont {Fortin}}, \bibinfo
  {author} {\bibfnamefont {H.}~\bibnamefont {Pais}}, \bibinfo {author}
  {\bibfnamefont {C.}~\bibnamefont {Provid\^encia}}, \bibinfo {author}
  {\bibfnamefont {A.~R.}\ \bibnamefont {Raduta}}, \ and\ \bibinfo {author}
  {\bibfnamefont {A.}~\bibnamefont {Sulaksono}},\ }\href {\doibase
  10.1103/PhysRevC.94.052801} {\bibfield  {journal} {\bibinfo  {journal} {Phys.
  Rev. C}\ }\textbf {\bibinfo {volume} {94}},\ \bibinfo {pages} {052801}
  (\bibinfo {year} {2016})}\BibitemShut {NoStop}%
\bibitem [{\citenamefont {Schneider}\ \emph {et~al.}(2019)\citenamefont
  {Schneider}, \citenamefont {Roberts}, \citenamefont {Ott},\ and\
  \citenamefont {O'Connor}}]{PhysRevC.100.055802}%
  \BibitemOpen
  \bibfield  {author} {\bibinfo {author} {\bibfnamefont {A.~S.}\ \bibnamefont
  {Schneider}}, \bibinfo {author} {\bibfnamefont {L.~F.}\ \bibnamefont
  {Roberts}}, \bibinfo {author} {\bibfnamefont {C.~D.}\ \bibnamefont {Ott}}, \
  and\ \bibinfo {author} {\bibfnamefont {E.}~\bibnamefont {O'Connor}},\ }\href
  {\doibase 10.1103/PhysRevC.100.055802} {\bibfield  {journal} {\bibinfo
  {journal} {Phys. Rev. C}\ }\textbf {\bibinfo {volume} {100}},\ \bibinfo
  {pages} {055802} (\bibinfo {year} {2019})}\BibitemShut {NoStop}%
\bibitem [{\citenamefont {Gil}\ \emph {et~al.}(2021)\citenamefont {Gil},
  \citenamefont {Kim}, \citenamefont {Papakonstantinou},\ and\ \citenamefont
  {Hyun}}]{PhysRevC.103.034330}%
  \BibitemOpen
  \bibfield  {author} {\bibinfo {author} {\bibfnamefont {H.}~\bibnamefont
  {Gil}}, \bibinfo {author} {\bibfnamefont {Y.-M.}\ \bibnamefont {Kim}},
  \bibinfo {author} {\bibfnamefont {P.}~\bibnamefont {Papakonstantinou}}, \
  and\ \bibinfo {author} {\bibfnamefont {C.~H.}\ \bibnamefont {Hyun}},\ }\href
  {\doibase 10.1103/PhysRevC.103.034330} {\bibfield  {journal} {\bibinfo
  {journal} {Phys. Rev. C}\ }\textbf {\bibinfo {volume} {103}},\ \bibinfo
  {pages} {034330} (\bibinfo {year} {2021})}\BibitemShut {NoStop}%
\bibitem [{\citenamefont {Zhang}\ \emph {et~al.}(2008)\citenamefont {Zhang},
  \citenamefont {Danielewicz}, \citenamefont {Famiano}, \citenamefont {Li},
  \citenamefont {Lynch},\ and\ \citenamefont {Tsang}}]{ZHANG2008145}%
  \BibitemOpen
  \bibfield  {author} {\bibinfo {author} {\bibfnamefont {Y.}~\bibnamefont
  {Zhang}}, \bibinfo {author} {\bibfnamefont {P.}~\bibnamefont {Danielewicz}},
  \bibinfo {author} {\bibfnamefont {M.}~\bibnamefont {Famiano}}, \bibinfo
  {author} {\bibfnamefont {Z.}~\bibnamefont {Li}}, \bibinfo {author}
  {\bibfnamefont {W.}~\bibnamefont {Lynch}}, \ and\ \bibinfo {author}
  {\bibfnamefont {M.}~\bibnamefont {Tsang}},\ }\href {\doibase
  https://doi.org/10.1016/j.physletb.2008.03.075} {\bibfield  {journal}
  {\bibinfo  {journal} {Physics Letters B}\ }\textbf {\bibinfo {volume}
  {664}},\ \bibinfo {pages} {145} (\bibinfo {year} {2008})}\BibitemShut
  {NoStop}%
\bibitem [{\citenamefont {Hagel}\ \emph {et~al.}(2014)\citenamefont {Hagel},
  \citenamefont {Natowitz},\ and\ \citenamefont {R\"opke}}]{Hagel:2014wja}%
  \BibitemOpen
  \bibfield  {author} {\bibinfo {author} {\bibfnamefont {K.}~\bibnamefont
  {Hagel}}, \bibinfo {author} {\bibfnamefont {J.~B.}\ \bibnamefont {Natowitz}},
  \ and\ \bibinfo {author} {\bibfnamefont {G.}~\bibnamefont {R\"opke}},\ }\href
  {\doibase 10.1140/epja/i2014-14039-4} {\bibfield  {journal} {\bibinfo
  {journal} {Eur. Phys. J. A}\ }\textbf {\bibinfo {volume} {50}},\ \bibinfo
  {pages} {39} (\bibinfo {year} {2014})},\ \Eprint
  {http://arxiv.org/abs/1401.2074} {arXiv:1401.2074 [nucl-ex]} \BibitemShut
  {NoStop}%
\bibitem [{\citenamefont {Wada}\ \emph {et~al.}(2012)\citenamefont {Wada},
  \citenamefont {Hagel}, \citenamefont {Qin}, \citenamefont {Natowitz},
  \citenamefont {Ma}, \citenamefont {R\"opke}, \citenamefont {Shlomo},
  \citenamefont {Bonasera}, \citenamefont {Typel}, \citenamefont {Chen},
  \citenamefont {Huang}, \citenamefont {Wang}, \citenamefont {Zheng},
  \citenamefont {Kowalski}, \citenamefont {Bottosso}, \citenamefont {Barbui},
  \citenamefont {Rodrigues}, \citenamefont {Schmidt}, \citenamefont {Fabris},
  \citenamefont {Lunardon}, \citenamefont {Moretto}, \citenamefont {Nebbia},
  \citenamefont {Pesente}, \citenamefont {Rizzi}, \citenamefont {Viesti},
  \citenamefont {Cinausero}, \citenamefont {Prete}, \citenamefont {Keutgen},
  \citenamefont {El~Masri},\ and\ \citenamefont {Majka}}]{PhysRevC.85.064618}%
  \BibitemOpen
  \bibfield  {author} {\bibinfo {author} {\bibfnamefont {R.}~\bibnamefont
  {Wada}}, \bibinfo {author} {\bibfnamefont {K.}~\bibnamefont {Hagel}},
  \bibinfo {author} {\bibfnamefont {L.}~\bibnamefont {Qin}}, \bibinfo {author}
  {\bibfnamefont {J.~B.}\ \bibnamefont {Natowitz}}, \bibinfo {author}
  {\bibfnamefont {Y.~G.}\ \bibnamefont {Ma}}, \bibinfo {author} {\bibfnamefont
  {G.}~\bibnamefont {R\"opke}}, \bibinfo {author} {\bibfnamefont
  {S.}~\bibnamefont {Shlomo}}, \bibinfo {author} {\bibfnamefont
  {A.}~\bibnamefont {Bonasera}}, \bibinfo {author} {\bibfnamefont
  {S.}~\bibnamefont {Typel}}, \bibinfo {author} {\bibfnamefont
  {Z.}~\bibnamefont {Chen}}, \bibinfo {author} {\bibfnamefont {M.}~\bibnamefont
  {Huang}}, \bibinfo {author} {\bibfnamefont {J.}~\bibnamefont {Wang}},
  \bibinfo {author} {\bibfnamefont {H.}~\bibnamefont {Zheng}}, \bibinfo
  {author} {\bibfnamefont {S.}~\bibnamefont {Kowalski}}, \bibinfo {author}
  {\bibfnamefont {C.}~\bibnamefont {Bottosso}}, \bibinfo {author}
  {\bibfnamefont {M.}~\bibnamefont {Barbui}}, \bibinfo {author} {\bibfnamefont
  {M.~R.~D.}\ \bibnamefont {Rodrigues}}, \bibinfo {author} {\bibfnamefont
  {K.}~\bibnamefont {Schmidt}}, \bibinfo {author} {\bibfnamefont
  {D.}~\bibnamefont {Fabris}}, \bibinfo {author} {\bibfnamefont
  {M.}~\bibnamefont {Lunardon}}, \bibinfo {author} {\bibfnamefont
  {S.}~\bibnamefont {Moretto}}, \bibinfo {author} {\bibfnamefont
  {G.}~\bibnamefont {Nebbia}}, \bibinfo {author} {\bibfnamefont
  {S.}~\bibnamefont {Pesente}}, \bibinfo {author} {\bibfnamefont
  {V.}~\bibnamefont {Rizzi}}, \bibinfo {author} {\bibfnamefont
  {G.}~\bibnamefont {Viesti}}, \bibinfo {author} {\bibfnamefont
  {M.}~\bibnamefont {Cinausero}}, \bibinfo {author} {\bibfnamefont
  {G.}~\bibnamefont {Prete}}, \bibinfo {author} {\bibfnamefont
  {T.}~\bibnamefont {Keutgen}}, \bibinfo {author} {\bibfnamefont
  {Y.}~\bibnamefont {El~Masri}}, \ and\ \bibinfo {author} {\bibfnamefont
  {Z.}~\bibnamefont {Majka}},\ }\href {\doibase 10.1103/PhysRevC.85.064618}
  {\bibfield  {journal} {\bibinfo  {journal} {Phys. Rev. C}\ }\textbf {\bibinfo
  {volume} {85}},\ \bibinfo {pages} {064618} (\bibinfo {year}
  {2012})}\BibitemShut {NoStop}%
\bibitem [{\citenamefont {Kumar}\ \emph {et~al.}(2020)\citenamefont {Kumar},
  \citenamefont {Das}, \citenamefont {Biswal}, \citenamefont {Kumar},\ and\
  \citenamefont {Patra}}]{Kumar_2020}%
  \BibitemOpen
  \bibfield  {author} {\bibinfo {author} {\bibfnamefont {A.}~\bibnamefont
  {Kumar}}, \bibinfo {author} {\bibfnamefont {H.~C.}\ \bibnamefont {Das}},
  \bibinfo {author} {\bibfnamefont {S.~K.}\ \bibnamefont {Biswal}}, \bibinfo
  {author} {\bibfnamefont {B.}~\bibnamefont {Kumar}}, \ and\ \bibinfo {author}
  {\bibfnamefont {S.~K.}\ \bibnamefont {Patra}},\ }\href {\doibase
  10.1140/epjc/s10052-020-8353-4} {\bibfield  {journal} {\bibinfo  {journal}
  {The European Physical Journal C}\ }\textbf {\bibinfo {volume} {80}}
  (\bibinfo {year} {2020}),\ 10.1140/epjc/s10052-020-8353-4}\BibitemShut
  {NoStop}%
\bibitem [{\citenamefont {Zhang}\ and\ \citenamefont
  {Chen}(2013)}]{ZHANG2013234}%
  \BibitemOpen
  \bibfield  {author} {\bibinfo {author} {\bibfnamefont {Z.}~\bibnamefont
  {Zhang}}\ and\ \bibinfo {author} {\bibfnamefont {L.-W.}\ \bibnamefont
  {Chen}},\ }\href {\doibase https://doi.org/10.1016/j.physletb.2013.08.002}
  {\bibfield  {journal} {\bibinfo  {journal} {Physics Letters B}\ }\textbf
  {\bibinfo {volume} {726}},\ \bibinfo {pages} {234} (\bibinfo {year}
  {2013})}\BibitemShut {NoStop}%
\bibitem [{\citenamefont {Trippa}\ \emph {et~al.}(2008)\citenamefont {Trippa},
  \citenamefont {Col\`o},\ and\ \citenamefont {Vigezzi}}]{PhysRevC.77.061304}%
  \BibitemOpen
  \bibfield  {author} {\bibinfo {author} {\bibfnamefont {L.}~\bibnamefont
  {Trippa}}, \bibinfo {author} {\bibfnamefont {G.}~\bibnamefont {Col\`o}}, \
  and\ \bibinfo {author} {\bibfnamefont {E.}~\bibnamefont {Vigezzi}},\ }\href
  {\doibase 10.1103/PhysRevC.77.061304} {\bibfield  {journal} {\bibinfo
  {journal} {Phys. Rev. C}\ }\textbf {\bibinfo {volume} {77}},\ \bibinfo
  {pages} {061304} (\bibinfo {year} {2008})}\BibitemShut {NoStop}%
\bibitem [{\citenamefont {Tsang}\ \emph {et~al.}(2009)\citenamefont {Tsang},
  \citenamefont {Zhang}, \citenamefont {Danielewicz}, \citenamefont {Famiano},
  \citenamefont {Li}, \citenamefont {Lynch},\ and\ \citenamefont
  {Steiner}}]{PhysRevLett.102.122701}%
  \BibitemOpen
  \bibfield  {author} {\bibinfo {author} {\bibfnamefont {M.~B.}\ \bibnamefont
  {Tsang}}, \bibinfo {author} {\bibfnamefont {Y.}~\bibnamefont {Zhang}},
  \bibinfo {author} {\bibfnamefont {P.}~\bibnamefont {Danielewicz}}, \bibinfo
  {author} {\bibfnamefont {M.}~\bibnamefont {Famiano}}, \bibinfo {author}
  {\bibfnamefont {Z.}~\bibnamefont {Li}}, \bibinfo {author} {\bibfnamefont
  {W.~G.}\ \bibnamefont {Lynch}}, \ and\ \bibinfo {author} {\bibfnamefont
  {A.~W.}\ \bibnamefont {Steiner}},\ }\href {\doibase
  10.1103/PhysRevLett.102.122701} {\bibfield  {journal} {\bibinfo  {journal}
  {Phys. Rev. Lett.}\ }\textbf {\bibinfo {volume} {102}},\ \bibinfo {pages}
  {122701} (\bibinfo {year} {2009})}\BibitemShut {NoStop}%
\bibitem [{\citenamefont {Olson}(2000)}]{PhysRevC.63.015802}%
  \BibitemOpen
  \bibfield  {author} {\bibinfo {author} {\bibfnamefont {T.~S.}\ \bibnamefont
  {Olson}},\ }\href {\doibase 10.1103/PhysRevC.63.015802} {\bibfield  {journal}
  {\bibinfo  {journal} {Phys. Rev. C}\ }\textbf {\bibinfo {volume} {63}},\
  \bibinfo {pages} {015802} (\bibinfo {year} {2000})}\BibitemShut {NoStop}%
\bibitem [{\citenamefont {{Hashimoto}}\ \emph {et~al.}(1994)\citenamefont
  {{Hashimoto}}, \citenamefont {{Oyamatsu}},\ and\ \citenamefont
  {{Eriguchi}}}]{1994ApJ...436..257H}%
  \BibitemOpen
  \bibfield  {author} {\bibinfo {author} {\bibfnamefont {M.-A.}\ \bibnamefont
  {{Hashimoto}}}, \bibinfo {author} {\bibfnamefont {K.}~\bibnamefont
  {{Oyamatsu}}}, \ and\ \bibinfo {author} {\bibfnamefont {Y.}~\bibnamefont
  {{Eriguchi}}},\ }\href {\doibase 10.1086/174899} {\bibfield  {journal}
  {\bibinfo  {journal} {apj}\ }\textbf {\bibinfo {volume} {436}},\ \bibinfo
  {pages} {257} (\bibinfo {year} {1994})}\BibitemShut {NoStop}%
\bibitem [{\citenamefont {Bertsch}\ \emph {et~al.}(1983)\citenamefont
  {Bertsch}, \citenamefont {Bortignon},\ and\ \citenamefont
  {Broglia}}]{RevModPhys.55.287}%
  \BibitemOpen
  \bibfield  {author} {\bibinfo {author} {\bibfnamefont {G.~F.}\ \bibnamefont
  {Bertsch}}, \bibinfo {author} {\bibfnamefont {P.~F.}\ \bibnamefont
  {Bortignon}}, \ and\ \bibinfo {author} {\bibfnamefont {R.~A.}\ \bibnamefont
  {Broglia}},\ }\href {\doibase 10.1103/RevModPhys.55.287} {\bibfield
  {journal} {\bibinfo  {journal} {Rev. Mod. Phys.}\ }\textbf {\bibinfo {volume}
  {55}},\ \bibinfo {pages} {287} (\bibinfo {year} {1983})}\BibitemShut
  {NoStop}%
\bibitem [{\citenamefont {Bertrand}(1981)}]{BERTRAND1981129}%
  \BibitemOpen
  \bibfield  {author} {\bibinfo {author} {\bibfnamefont {F.~E.}\ \bibnamefont
  {Bertrand}},\ }\href {\doibase https://doi.org/10.1016/0375-9474(81)90596-0}
  {\bibfield  {journal} {\bibinfo  {journal} {Nuclear Physics A}\ }\textbf
  {\bibinfo {volume} {354}},\ \bibinfo {pages} {129} (\bibinfo {year}
  {1981})}\BibitemShut {NoStop}%
\bibitem [{\citenamefont {Youngblood}\ \emph {et~al.}(2004)\citenamefont
  {Youngblood}, \citenamefont {Lui}, \citenamefont {Clark}, \citenamefont
  {John}, \citenamefont {Tokimoto},\ and\ \citenamefont
  {Chen}}]{PhysRevC.69.034315}%
  \BibitemOpen
  \bibfield  {author} {\bibinfo {author} {\bibfnamefont {D.~H.}\ \bibnamefont
  {Youngblood}}, \bibinfo {author} {\bibfnamefont {Y.-W.}\ \bibnamefont {Lui}},
  \bibinfo {author} {\bibfnamefont {H.~L.}\ \bibnamefont {Clark}}, \bibinfo
  {author} {\bibfnamefont {B.}~\bibnamefont {John}}, \bibinfo {author}
  {\bibfnamefont {Y.}~\bibnamefont {Tokimoto}}, \ and\ \bibinfo {author}
  {\bibfnamefont {X.}~\bibnamefont {Chen}},\ }\href {\doibase
  10.1103/PhysRevC.69.034315} {\bibfield  {journal} {\bibinfo  {journal} {Phys.
  Rev. C}\ }\textbf {\bibinfo {volume} {69}},\ \bibinfo {pages} {034315}
  (\bibinfo {year} {2004})}\BibitemShut {NoStop}%
\bibitem [{\citenamefont {{Buenerd, M.}}(1984)}]{Buenerd}%
  \BibitemOpen
  \bibfield  {author} {\bibinfo {author} {\bibnamefont {{Buenerd, M.}}},\
  }\href {\doibase 10.1051/jphyscol:1984411} {\bibfield  {journal} {\bibinfo
  {journal} {J. Phys. Colloques}\ }\textbf {\bibinfo {volume} {45}},\ \bibinfo
  {pages} {C4} (\bibinfo {year} {1984})}\BibitemShut {NoStop}%
\bibitem [{\citenamefont {Centelles}\ \emph {et~al.}(2005)\citenamefont
  {Centelles}, \citenamefont {Vi\~nas}, \citenamefont {Patra}, \citenamefont
  {De},\ and\ \citenamefont {Sil}}]{PhysRevC.72.014304}%
  \BibitemOpen
  \bibfield  {author} {\bibinfo {author} {\bibfnamefont {M.}~\bibnamefont
  {Centelles}}, \bibinfo {author} {\bibfnamefont {X.}~\bibnamefont {Vi\~nas}},
  \bibinfo {author} {\bibfnamefont {S.~K.}\ \bibnamefont {Patra}}, \bibinfo
  {author} {\bibfnamefont {J.~N.}\ \bibnamefont {De}}, \ and\ \bibinfo {author}
  {\bibfnamefont {T.}~\bibnamefont {Sil}},\ }\href {\doibase
  10.1103/PhysRevC.72.014304} {\bibfield  {journal} {\bibinfo  {journal} {Phys.
  Rev. C}\ }\textbf {\bibinfo {volume} {72}},\ \bibinfo {pages} {014304}
  (\bibinfo {year} {2005})}\BibitemShut {NoStop}%
\bibitem [{\citenamefont {Patra}\ \emph
  {et~al.}(2002{\natexlab{a}})\citenamefont {Patra}, \citenamefont {Viñas},
  \citenamefont {Centelles},\ and\ \citenamefont {Del~Estal}}]{Patra_2002}%
  \BibitemOpen
  \bibfield  {author} {\bibinfo {author} {\bibfnamefont {S.}~\bibnamefont
  {Patra}}, \bibinfo {author} {\bibfnamefont {X.}~\bibnamefont {Viñas}},
  \bibinfo {author} {\bibfnamefont {M.}~\bibnamefont {Centelles}}, \ and\
  \bibinfo {author} {\bibfnamefont {M.}~\bibnamefont {Del~Estal}},\ }\href
  {\doibase 10.1016/s0375-9474(01)01531-7} {\bibfield  {journal} {\bibinfo
  {journal} {Nuclear Physics A}\ }\textbf {\bibinfo {volume} {703}},\ \bibinfo
  {pages} {240–268} (\bibinfo {year} {2002}{\natexlab{a}})}\BibitemShut
  {NoStop}%
\bibitem [{\citenamefont {Bednarek}\ \emph {et~al.}(2020)\citenamefont
  {Bednarek}, \citenamefont {Sładkowski},\ and\ \citenamefont
  {Syska}}]{sym12060898}%
  \BibitemOpen
  \bibfield  {author} {\bibinfo {author} {\bibfnamefont {I.}~\bibnamefont
  {Bednarek}}, \bibinfo {author} {\bibfnamefont {J.}~\bibnamefont
  {Sładkowski}}, \ and\ \bibinfo {author} {\bibfnamefont {J.}~\bibnamefont
  {Syska}},\ }\href {\doibase 10.3390/sym12060898} {\bibfield  {journal}
  {\bibinfo  {journal} {Symmetry}\ }\textbf {\bibinfo {volume} {12}} (\bibinfo
  {year} {2020}),\ 10.3390/sym12060898}\BibitemShut {NoStop}%
\bibitem [{\citenamefont {{Lattimer}}(2012)}]{2012ARNPS..62..485L}%
  \BibitemOpen
  \bibfield  {author} {\bibinfo {author} {\bibfnamefont {J.~M.}\ \bibnamefont
  {{Lattimer}}},\ }\href {\doibase 10.1146/annurev-nucl-102711-095018}
  {\bibfield  {journal} {\bibinfo  {journal} {Annual Review of Nuclear and
  Particle Science}\ }\textbf {\bibinfo {volume} {62}},\ \bibinfo {pages} {485}
  (\bibinfo {year} {2012})},\ \Eprint {http://arxiv.org/abs/1305.3510}
  {arXiv:1305.3510 [nucl-th]} \BibitemShut {NoStop}%
\bibitem [{\citenamefont {Engvik}\ \emph {et~al.}(1994)\citenamefont {Engvik},
  \citenamefont {Hjorth-Jensen}, \citenamefont {Osnes}, \citenamefont {Bao},\
  and\ \citenamefont {\O{}stgaard}}]{PhysRevLett.73.2650}%
  \BibitemOpen
  \bibfield  {author} {\bibinfo {author} {\bibfnamefont {L.}~\bibnamefont
  {Engvik}}, \bibinfo {author} {\bibfnamefont {M.}~\bibnamefont
  {Hjorth-Jensen}}, \bibinfo {author} {\bibfnamefont {E.}~\bibnamefont
  {Osnes}}, \bibinfo {author} {\bibfnamefont {G.}~\bibnamefont {Bao}}, \ and\
  \bibinfo {author} {\bibfnamefont {E.}~\bibnamefont {\O{}stgaard}},\ }\href
  {\doibase 10.1103/PhysRevLett.73.2650} {\bibfield  {journal} {\bibinfo
  {journal} {Phys. Rev. Lett.}\ }\textbf {\bibinfo {volume} {73}},\ \bibinfo
  {pages} {2650} (\bibinfo {year} {1994})}\BibitemShut {NoStop}%
\bibitem [{\citenamefont {Bombaci}\ and\ \citenamefont
  {Logoteta}(2018)}]{Bombaci_2018}%
  \BibitemOpen
  \bibfield  {author} {\bibinfo {author} {\bibfnamefont {I.}~\bibnamefont
  {Bombaci}}\ and\ \bibinfo {author} {\bibfnamefont {D.}~\bibnamefont
  {Logoteta}},\ }\href {\doibase 10.1051/0004-6361/201731604} {\bibfield
  {journal} {\bibinfo  {journal} {Astronomy \& Astrophysics}\ }\textbf
  {\bibinfo {volume} {609}},\ \bibinfo {pages} {A128} (\bibinfo {year}
  {2018})}\BibitemShut {NoStop}%
\bibitem [{\citenamefont {Abbott}\ \emph {et~al.}(2017)\citenamefont {Abbott}
  \emph {et~al.}}]{PhysRevLett.119.161101}%
  \BibitemOpen
  \bibfield  {author} {\bibinfo {author} {\bibfnamefont {B.~P.}\ \bibnamefont
  {Abbott}} \emph {et~al.} (\bibinfo {collaboration} {The LIGO Scientific
  Collaboration and the Virgo Collaboration}),\ }\href {\doibase
  10.1103/PhysRevLett.119.161101} {\bibfield  {journal} {\bibinfo  {journal}
  {Phys. Rev. Lett.}\ }\textbf {\bibinfo {volume} {119}},\ \bibinfo {pages}
  {161101} (\bibinfo {year} {2017})}\BibitemShut {NoStop}%
\bibitem [{\citenamefont {Boguta}(1981)}]{BOGUTA1981255}%
  \BibitemOpen
  \bibfield  {author} {\bibinfo {author} {\bibfnamefont {J.}~\bibnamefont
  {Boguta}},\ }\href {\doibase https://doi.org/10.1016/0370-2693(81)90529-3}
  {\bibfield  {journal} {\bibinfo  {journal} {Physics Letters B}\ }\textbf
  {\bibinfo {volume} {106}},\ \bibinfo {pages} {255} (\bibinfo {year}
  {1981})}\BibitemShut {NoStop}%
\bibitem [{\citenamefont {Observatory}()}]{NASA}%
  \BibitemOpen
  \bibfield  {author} {\bibinfo {author} {\bibfnamefont {C.~X.-R.}\
  \bibnamefont {Observatory}},\ }\href
  {https://www.nasa.gov/mission_pages/chandra/main/index.html} {}\BibitemShut
  {NoStop}%
\bibitem [{\citenamefont {Greif}\ \emph {et~al.}(2020)\citenamefont {Greif},
  \citenamefont {Hebeler}, \citenamefont {Lattimer}, \citenamefont {Pethick},\
  and\ \citenamefont {Schwenk}}]{Greif_2020}%
  \BibitemOpen
  \bibfield  {author} {\bibinfo {author} {\bibfnamefont {S.~K.}\ \bibnamefont
  {Greif}}, \bibinfo {author} {\bibfnamefont {K.}~\bibnamefont {Hebeler}},
  \bibinfo {author} {\bibfnamefont {J.~M.}\ \bibnamefont {Lattimer}}, \bibinfo
  {author} {\bibfnamefont {C.~J.}\ \bibnamefont {Pethick}}, \ and\ \bibinfo
  {author} {\bibfnamefont {A.}~\bibnamefont {Schwenk}},\ }\href {\doibase
  10.3847/1538-4357/abaf55} {\bibfield  {journal} {\bibinfo  {journal} {The
  Astrophysical Journal}\ }\textbf {\bibinfo {volume} {901}},\ \bibinfo {pages}
  {155} (\bibinfo {year} {2020})}\BibitemShut {NoStop}%
\bibitem [{\citenamefont {Skyrme}\ and\ \citenamefont
  {Schonland}(1961)}]{doi:10.1098/rspa.1961.0018}%
  \BibitemOpen
  \bibfield  {author} {\bibinfo {author} {\bibfnamefont {T.~H.~R.}\
  \bibnamefont {Skyrme}}\ and\ \bibinfo {author} {\bibfnamefont {B.~F.~J.}\
  \bibnamefont {Schonland}},\ }\href {\doibase 10.1098/rspa.1961.0018}
  {\bibfield  {journal} {\bibinfo  {journal} {Proceedings of the Royal Society
  of London. Series A. Mathematical and Physical Sciences}\ }\textbf {\bibinfo
  {volume} {260}},\ \bibinfo {pages} {127} (\bibinfo {year}
  {1961})}\BibitemShut {NoStop}%
\bibitem [{\citenamefont {Decharg\'e}\ and\ \citenamefont
  {Gogny}(1980)}]{PhysRevC.21.1568}%
  \BibitemOpen
  \bibfield  {author} {\bibinfo {author} {\bibfnamefont {J.}~\bibnamefont
  {Decharg\'e}}\ and\ \bibinfo {author} {\bibfnamefont {D.}~\bibnamefont
  {Gogny}},\ }\href {\doibase 10.1103/PhysRevC.21.1568} {\bibfield  {journal}
  {\bibinfo  {journal} {Phys. Rev. C}\ }\textbf {\bibinfo {volume} {21}},\
  \bibinfo {pages} {1568} (\bibinfo {year} {1980})}\BibitemShut {NoStop}%
\bibitem [{\citenamefont {Walecka}(1974)}]{WALECKA1974491}%
  \BibitemOpen
  \bibfield  {author} {\bibinfo {author} {\bibfnamefont {J.}~\bibnamefont
  {Walecka}},\ }\href {\doibase https://doi.org/10.1016/0003-4916(74)90208-5}
  {\bibfield  {journal} {\bibinfo  {journal} {Annals of Physics}\ }\textbf
  {\bibinfo {volume} {83}},\ \bibinfo {pages} {491} (\bibinfo {year}
  {1974})}\BibitemShut {NoStop}%
\bibitem [{\citenamefont {Gambhir}\ \emph {et~al.}(1990)\citenamefont
  {Gambhir}, \citenamefont {Ring},\ and\ \citenamefont
  {Thimet}}]{GAMBHIR1990132}%
  \BibitemOpen
  \bibfield  {author} {\bibinfo {author} {\bibfnamefont {Y.}~\bibnamefont
  {Gambhir}}, \bibinfo {author} {\bibfnamefont {P.}~\bibnamefont {Ring}}, \
  and\ \bibinfo {author} {\bibfnamefont {A.}~\bibnamefont {Thimet}},\ }\href
  {\doibase https://doi.org/10.1016/0003-4916(90)90330-Q} {\bibfield  {journal}
  {\bibinfo  {journal} {Annals of Physics}\ }\textbf {\bibinfo {volume}
  {198}},\ \bibinfo {pages} {132} (\bibinfo {year} {1990})}\BibitemShut
  {NoStop}%
\bibitem [{\citenamefont {Adam}\ \emph {et~al.}(2020)\citenamefont {Adam},
  \citenamefont {Martín-Caro}, \citenamefont {Huidobro}, \citenamefont
  {Vázquez},\ and\ \citenamefont {Wereszczynski}}]{ADAM2020135928}%
  \BibitemOpen
  \bibfield  {author} {\bibinfo {author} {\bibfnamefont {C.}~\bibnamefont
  {Adam}}, \bibinfo {author} {\bibfnamefont {A.~G.}\ \bibnamefont
  {Martín-Caro}}, \bibinfo {author} {\bibfnamefont {M.}~\bibnamefont
  {Huidobro}}, \bibinfo {author} {\bibfnamefont {R.}~\bibnamefont {Vázquez}},
  \ and\ \bibinfo {author} {\bibfnamefont {A.}~\bibnamefont {Wereszczynski}},\
  }\href {\doibase https://doi.org/10.1016/j.physletb.2020.135928} {\bibfield
  {journal} {\bibinfo  {journal} {Physics Letters B}\ }\textbf {\bibinfo
  {volume} {811}},\ \bibinfo {pages} {135928} (\bibinfo {year}
  {2020})}\BibitemShut {NoStop}%
\bibitem [{\citenamefont {Adam}\ \emph {et~al.}(2015)\citenamefont {Adam},
  \citenamefont {Haberichter},\ and\ \citenamefont
  {Wereszczynski}}]{Adam_2015}%
  \BibitemOpen
  \bibfield  {author} {\bibinfo {author} {\bibfnamefont {C.}~\bibnamefont
  {Adam}}, \bibinfo {author} {\bibfnamefont {M.}~\bibnamefont {Haberichter}}, \
  and\ \bibinfo {author} {\bibfnamefont {A.}~\bibnamefont {Wereszczynski}},\
  }\href {\doibase 10.1103/physrevc.92.055807} {\bibfield  {journal} {\bibinfo
  {journal} {Physical Review C}\ }\textbf {\bibinfo {volume} {92}} (\bibinfo
  {year} {2015}),\ 10.1103/physrevc.92.055807}\BibitemShut {NoStop}%
\bibitem [{\citenamefont {Kumar}\ \emph
  {et~al.}(2018{\natexlab{a}})\citenamefont {Kumar}, \citenamefont {Patra},\
  and\ \citenamefont {Agrawal}}]{BharatKumar}%
  \BibitemOpen
  \bibfield  {author} {\bibinfo {author} {\bibfnamefont {B.}~\bibnamefont
  {Kumar}}, \bibinfo {author} {\bibfnamefont {S.~K.}\ \bibnamefont {Patra}}, \
  and\ \bibinfo {author} {\bibfnamefont {B.~K.}\ \bibnamefont {Agrawal}},\
  }\href {\doibase 10.1103/PhysRevC.97.045806} {\bibfield  {journal} {\bibinfo
  {journal} {Phys. Rev. C}\ }\textbf {\bibinfo {volume} {97}},\ \bibinfo
  {pages} {045806} (\bibinfo {year} {2018}{\natexlab{a}})}\BibitemShut
  {NoStop}%
\bibitem [{\citenamefont {Kumar}\ \emph {et~al.}(2017)\citenamefont {Kumar},
  \citenamefont {Singh}, \citenamefont {Agrawal},\ and\ \citenamefont
  {Patra}}]{KUMAR2017197}%
  \BibitemOpen
  \bibfield  {author} {\bibinfo {author} {\bibfnamefont {B.}~\bibnamefont
  {Kumar}}, \bibinfo {author} {\bibfnamefont {S.}~\bibnamefont {Singh}},
  \bibinfo {author} {\bibfnamefont {B.}~\bibnamefont {Agrawal}}, \ and\
  \bibinfo {author} {\bibfnamefont {S.}~\bibnamefont {Patra}},\ }\href
  {\doibase https://doi.org/10.1016/j.nuclphysa.2017.07.001} {\bibfield
  {journal} {\bibinfo  {journal} {Nuclear Physics A}\ }\textbf {\bibinfo
  {volume} {966}},\ \bibinfo {pages} {197} (\bibinfo {year}
  {2017})}\BibitemShut {NoStop}%
\bibitem [{\citenamefont {Avancini}\ \emph {et~al.}(2003)\citenamefont
  {Avancini}, \citenamefont {Bracco}, \citenamefont {Chiapparini},\ and\
  \citenamefont {Menezes}}]{Avancini:2002kf}%
  \BibitemOpen
  \bibfield  {author} {\bibinfo {author} {\bibfnamefont {S.~S.}\ \bibnamefont
  {Avancini}}, \bibinfo {author} {\bibfnamefont {M.~E.}\ \bibnamefont
  {Bracco}}, \bibinfo {author} {\bibfnamefont {M.}~\bibnamefont {Chiapparini}},
  \ and\ \bibinfo {author} {\bibfnamefont {D.~P.}\ \bibnamefont {Menezes}},\
  }\href {\doibase 10.1103/PhysRevC.67.024301} {\bibfield  {journal} {\bibinfo
  {journal} {Phys. Rev. C}\ }\textbf {\bibinfo {volume} {67}},\ \bibinfo
  {pages} {024301} (\bibinfo {year} {2003})},\ \Eprint
  {http://arxiv.org/abs/nucl-th/0212080} {arXiv:nucl-th/0212080} \BibitemShut
  {NoStop}%
\bibitem [{\citenamefont {Singh}\ \emph {et~al.}(2014)\citenamefont {Singh},
  \citenamefont {Biswal}, \citenamefont {Bhuyan},\ and\ \citenamefont
  {Patra}}]{PhysRevC.89.044001}%
  \BibitemOpen
  \bibfield  {author} {\bibinfo {author} {\bibfnamefont {S.~K.}\ \bibnamefont
  {Singh}}, \bibinfo {author} {\bibfnamefont {S.~K.}\ \bibnamefont {Biswal}},
  \bibinfo {author} {\bibfnamefont {M.}~\bibnamefont {Bhuyan}}, \ and\ \bibinfo
  {author} {\bibfnamefont {S.~K.}\ \bibnamefont {Patra}},\ }\href {\doibase
  10.1103/PhysRevC.89.044001} {\bibfield  {journal} {\bibinfo  {journal} {Phys.
  Rev. C}\ }\textbf {\bibinfo {volume} {89}},\ \bibinfo {pages} {044001}
  (\bibinfo {year} {2014})}\BibitemShut {NoStop}%
\bibitem [{\citenamefont {Lalazissis}\ \emph {et~al.}(1997)\citenamefont
  {Lalazissis}, \citenamefont {K\"onig},\ and\ \citenamefont
  {Ring}}]{PhysRevC.55.540}%
  \BibitemOpen
  \bibfield  {author} {\bibinfo {author} {\bibfnamefont {G.~A.}\ \bibnamefont
  {Lalazissis}}, \bibinfo {author} {\bibfnamefont {J.}~\bibnamefont {K\"onig}},
  \ and\ \bibinfo {author} {\bibfnamefont {P.}~\bibnamefont {Ring}},\ }\href
  {\doibase 10.1103/PhysRevC.55.540} {\bibfield  {journal} {\bibinfo  {journal}
  {Phys. Rev. C}\ }\textbf {\bibinfo {volume} {55}},\ \bibinfo {pages} {540}
  (\bibinfo {year} {1997})}\BibitemShut {NoStop}%
\bibitem [{\citenamefont {Carbone}\ and\ \citenamefont
  {Schwenk}(2019)}]{PhysRevC.100.025805}%
  \BibitemOpen
  \bibfield  {author} {\bibinfo {author} {\bibfnamefont {A.}~\bibnamefont
  {Carbone}}\ and\ \bibinfo {author} {\bibfnamefont {A.}~\bibnamefont
  {Schwenk}},\ }\href {\doibase 10.1103/PhysRevC.100.025805} {\bibfield
  {journal} {\bibinfo  {journal} {Phys. Rev. C}\ }\textbf {\bibinfo {volume}
  {100}},\ \bibinfo {pages} {025805} (\bibinfo {year} {2019})}\BibitemShut
  {NoStop}%
\bibitem [{\citenamefont {Das}\ \emph {et~al.}(2020{\natexlab{a}})\citenamefont
  {Das}, \citenamefont {Kumar}, \citenamefont {Kumar}, \citenamefont {Biswal},
  \citenamefont {Nakatsukasa}, \citenamefont {Li},\ and\ \citenamefont
  {Patra}}]{HCDas}%
  \BibitemOpen
  \bibfield  {author} {\bibinfo {author} {\bibfnamefont {H.~C.}\ \bibnamefont
  {Das}}, \bibinfo {author} {\bibfnamefont {A.}~\bibnamefont {Kumar}}, \bibinfo
  {author} {\bibfnamefont {B.}~\bibnamefont {Kumar}}, \bibinfo {author}
  {\bibfnamefont {S.~K.}\ \bibnamefont {Biswal}}, \bibinfo {author}
  {\bibfnamefont {T.}~\bibnamefont {Nakatsukasa}}, \bibinfo {author}
  {\bibfnamefont {A.}~\bibnamefont {Li}}, \ and\ \bibinfo {author}
  {\bibfnamefont {S.~K.}\ \bibnamefont {Patra}},\ }\href {\doibase
  10.1093/mnras/staa1435} {\bibfield  {journal} {\bibinfo  {journal} {Monthly
  Notices of the Royal Astronomical Society}\ }\textbf {\bibinfo {volume}
  {495}},\ \bibinfo {pages} {4893} (\bibinfo {year}
  {2020}{\natexlab{a}})}\BibitemShut {NoStop}%
\bibitem [{\citenamefont {Garg}\ and\ \citenamefont
  {Colò}(2018)}]{GARG201855}%
  \BibitemOpen
  \bibfield  {author} {\bibinfo {author} {\bibfnamefont {U.}~\bibnamefont
  {Garg}}\ and\ \bibinfo {author} {\bibfnamefont {G.}~\bibnamefont {Colò}},\
  }\href {\doibase https://doi.org/10.1016/j.ppnp.2018.03.001} {\bibfield
  {journal} {\bibinfo  {journal} {Progress in Particle and Nuclear Physics}\
  }\textbf {\bibinfo {volume} {101}},\ \bibinfo {pages} {55} (\bibinfo {year}
  {2018})}\BibitemShut {NoStop}%
\bibitem [{\citenamefont {Danielewicz}\ and\ \citenamefont
  {Lee}(2014)}]{DANIELEWICZ20141}%
  \BibitemOpen
  \bibfield  {author} {\bibinfo {author} {\bibfnamefont {P.}~\bibnamefont
  {Danielewicz}}\ and\ \bibinfo {author} {\bibfnamefont {J.}~\bibnamefont
  {Lee}},\ }\href {\doibase https://doi.org/10.1016/j.nuclphysa.2013.11.005}
  {\bibfield  {journal} {\bibinfo  {journal} {Nuclear Physics A}\ }\textbf
  {\bibinfo {volume} {922}},\ \bibinfo {pages} {1} (\bibinfo {year}
  {2014})}\BibitemShut {NoStop}%
\bibitem [{\citenamefont {Das}\ \emph {et~al.}(2021{\natexlab{a}})\citenamefont
  {Das}, \citenamefont {Kumar},\ and\ \citenamefont {Patra}}]{DasPRD_2021}%
  \BibitemOpen
  \bibfield  {author} {\bibinfo {author} {\bibfnamefont {H.~C.}\ \bibnamefont
  {Das}}, \bibinfo {author} {\bibfnamefont {A.}~\bibnamefont {Kumar}}, \ and\
  \bibinfo {author} {\bibfnamefont {S.~K.}\ \bibnamefont {Patra}},\ }\href
  {\doibase 10.1103/PhysRevD.104.063028} {\bibfield  {journal} {\bibinfo
  {journal} {Phys. Rev. D}\ }\textbf {\bibinfo {volume} {104}},\ \bibinfo
  {pages} {063028} (\bibinfo {year} {2021}{\natexlab{a}})},\ \Eprint
  {http://arxiv.org/abs/2109.01853} {arXiv:2109.01853 [astro-ph.HE]}
  \BibitemShut {NoStop}%
\bibitem [{\citenamefont {Zyla}\ \emph {et~al.}(2020)\citenamefont {Zyla},
  \citenamefont {Barnett}, \citenamefont {Beringer},\ and\ \citenamefont
  {et~al.}}]{Zyla_2020}%
  \BibitemOpen
  \bibfield  {author} {\bibinfo {author} {\bibfnamefont {P.}~\bibnamefont
  {Zyla}}, \bibinfo {author} {\bibfnamefont {R.}~\bibnamefont {Barnett}},
  \bibinfo {author} {\bibfnamefont {J.}~\bibnamefont {Beringer}}, \ and\
  \bibinfo {author} {\bibnamefont {et~al.}},\ }\href
  {http://pdg.lbl.gov/2020/html/authors_2020.html} {\enquote {\bibinfo {title}
  {Particle data group},}\ } (\bibinfo {year} {2020})\BibitemShut {NoStop}%
\bibitem [{\citenamefont
  {Bethe}(1971)}]{doi:10.1146/annurev.ns.21.120171.000521}%
  \BibitemOpen
  \bibfield  {author} {\bibinfo {author} {\bibfnamefont {H.~A.}\ \bibnamefont
  {Bethe}},\ }\href {\doibase 10.1146/annurev.ns.21.120171.000521} {\bibfield
  {journal} {\bibinfo  {journal} {Annual Review of Nuclear Science}\ }\textbf
  {\bibinfo {volume} {21}},\ \bibinfo {pages} {93} (\bibinfo {year} {1971})},\
  \Eprint
  {http://arxiv.org/abs/https://doi.org/10.1146/annurev.ns.21.120171.000521}
  {https://doi.org/10.1146/annurev.ns.21.120171.000521} \BibitemShut {NoStop}%
\bibitem [{\citenamefont {Zimmerman}\ \emph {et~al.}(2020)\citenamefont
  {Zimmerman}, \citenamefont {Carson}, \citenamefont {Schumacher},
  \citenamefont {Steiner},\ and\ \citenamefont
  {Yagi}}]{zimmerman2020measuring}%
  \BibitemOpen
  \bibfield  {author} {\bibinfo {author} {\bibfnamefont {J.}~\bibnamefont
  {Zimmerman}}, \bibinfo {author} {\bibfnamefont {Z.}~\bibnamefont {Carson}},
  \bibinfo {author} {\bibfnamefont {K.}~\bibnamefont {Schumacher}}, \bibinfo
  {author} {\bibfnamefont {A.~W.}\ \bibnamefont {Steiner}}, \ and\ \bibinfo
  {author} {\bibfnamefont {K.}~\bibnamefont {Yagi}},\ }\href@noop {} {}
  (\bibinfo {year} {2020}),\ \Eprint {http://arxiv.org/abs/2002.03210}
  {arXiv:2002.03210 [astro-ph.HE]} \BibitemShut {NoStop}%
\bibitem [{\citenamefont {Brueckner}\ \emph
  {et~al.}(1968{\natexlab{a}})\citenamefont {Brueckner}, \citenamefont {Coon},\
  and\ \citenamefont {Dabrowski}}]{PhysRev.168.1184}%
  \BibitemOpen
  \bibfield  {author} {\bibinfo {author} {\bibfnamefont {K.~A.}\ \bibnamefont
  {Brueckner}}, \bibinfo {author} {\bibfnamefont {S.~A.}\ \bibnamefont {Coon}},
  \ and\ \bibinfo {author} {\bibfnamefont {J.}~\bibnamefont {Dabrowski}},\
  }\href {\doibase 10.1103/PhysRev.168.1184} {\bibfield  {journal} {\bibinfo
  {journal} {Phys. Rev.}\ }\textbf {\bibinfo {volume} {168}},\ \bibinfo {pages}
  {1184} (\bibinfo {year} {1968}{\natexlab{a}})}\BibitemShut {NoStop}%
\bibitem [{\citenamefont {Brueckner}\ \emph
  {et~al.}(1968{\natexlab{b}})\citenamefont {Brueckner}, \citenamefont
  {Buchler}, \citenamefont {Jorna},\ and\ \citenamefont
  {Lombard}}]{PhysRev.171.1188}%
  \BibitemOpen
  \bibfield  {author} {\bibinfo {author} {\bibfnamefont {K.~A.}\ \bibnamefont
  {Brueckner}}, \bibinfo {author} {\bibfnamefont {J.~R.}\ \bibnamefont
  {Buchler}}, \bibinfo {author} {\bibfnamefont {S.}~\bibnamefont {Jorna}}, \
  and\ \bibinfo {author} {\bibfnamefont {R.~J.}\ \bibnamefont {Lombard}},\
  }\href {\doibase 10.1103/PhysRev.171.1188} {\bibfield  {journal} {\bibinfo
  {journal} {Phys. Rev.}\ }\textbf {\bibinfo {volume} {171}},\ \bibinfo {pages}
  {1188} (\bibinfo {year} {1968}{\natexlab{b}})}\BibitemShut {NoStop}%
\bibitem [{\citenamefont {Coester}\ \emph {et~al.}(1972)\citenamefont
  {Coester}, \citenamefont {Day},\ and\ \citenamefont
  {Goodman}}]{PhysRevC.5.1135}%
  \BibitemOpen
  \bibfield  {author} {\bibinfo {author} {\bibfnamefont {F.}~\bibnamefont
  {Coester}}, \bibinfo {author} {\bibfnamefont {B.}~\bibnamefont {Day}}, \ and\
  \bibinfo {author} {\bibfnamefont {A.}~\bibnamefont {Goodman}},\ }\href
  {\doibase 10.1103/PhysRevC.5.1135} {\bibfield  {journal} {\bibinfo  {journal}
  {Phys. Rev. C}\ }\textbf {\bibinfo {volume} {5}},\ \bibinfo {pages} {1135}
  (\bibinfo {year} {1972})}\BibitemShut {NoStop}%
\bibitem [{\citenamefont {Kumar}\ \emph {et~al.}(2021)\citenamefont {Kumar},
  \citenamefont {Das}, \citenamefont {Kaur}, \citenamefont {Bhuyan},\ and\
  \citenamefont {Patra}}]{PhysRevC.103.024305}%
  \BibitemOpen
  \bibfield  {author} {\bibinfo {author} {\bibfnamefont {A.}~\bibnamefont
  {Kumar}}, \bibinfo {author} {\bibfnamefont {H.~C.}\ \bibnamefont {Das}},
  \bibinfo {author} {\bibfnamefont {M.}~\bibnamefont {Kaur}}, \bibinfo {author}
  {\bibfnamefont {M.}~\bibnamefont {Bhuyan}}, \ and\ \bibinfo {author}
  {\bibfnamefont {S.~K.}\ \bibnamefont {Patra}},\ }\href {\doibase
  10.1103/PhysRevC.103.024305} {\bibfield  {journal} {\bibinfo  {journal}
  {Phys. Rev. C}\ }\textbf {\bibinfo {volume} {103}},\ \bibinfo {pages}
  {024305} (\bibinfo {year} {2021})}\BibitemShut {NoStop}%
\bibitem [{\citenamefont {Antonov}\ \emph {et~al.}(1980)\citenamefont
  {Antonov}, \citenamefont {Nikolaev},\ and\ \citenamefont
  {Petkov}}]{Antonov1980}%
  \BibitemOpen
  \bibfield  {author} {\bibinfo {author} {\bibfnamefont {A.}~\bibnamefont
  {Antonov}}, \bibinfo {author} {\bibfnamefont {V.}~\bibnamefont {Nikolaev}}, \
  and\ \bibinfo {author} {\bibfnamefont {I.}~\bibnamefont {Petkov}},\ }\href
  {\doibase 10.1007/BF01892806} {\bibfield  {journal} {\bibinfo  {journal}
  {Zeitschrift für Physik A Atoms and Nuclei}\ }\textbf {\bibinfo {volume}
  {297}} (\bibinfo {year} {1980}),\ 10.1007/BF01892806}\BibitemShut {NoStop}%
\bibitem [{\citenamefont {Antonov}\ \emph {et~al.}(1994)\citenamefont
  {Antonov}, \citenamefont {Kadrev},\ and\ \citenamefont
  {Hodgson}}]{PhysRevC.50.164}%
  \BibitemOpen
  \bibfield  {author} {\bibinfo {author} {\bibfnamefont {A.~N.}\ \bibnamefont
  {Antonov}}, \bibinfo {author} {\bibfnamefont {D.~N.}\ \bibnamefont {Kadrev}},
  \ and\ \bibinfo {author} {\bibfnamefont {P.~E.}\ \bibnamefont {Hodgson}},\
  }\href {\doibase 10.1103/PhysRevC.50.164} {\bibfield  {journal} {\bibinfo
  {journal} {Phys. Rev. C}\ }\textbf {\bibinfo {volume} {50}},\ \bibinfo
  {pages} {164} (\bibinfo {year} {1994})}\BibitemShut {NoStop}%
\bibitem [{\citenamefont {Gaidarov}\ \emph {et~al.}(2020)\citenamefont
  {Gaidarov}, \citenamefont {Moumene}, \citenamefont {Antonov}, \citenamefont
  {Kadrev}, \citenamefont {Sarriguren},\ and\ \citenamefont
  {Guerra}}]{Gaidarov2020ProtonAN}%
  \BibitemOpen
  \bibfield  {author} {\bibinfo {author} {\bibfnamefont {M.}~\bibnamefont
  {Gaidarov}}, \bibinfo {author} {\bibfnamefont {I.}~\bibnamefont {Moumene}},
  \bibinfo {author} {\bibfnamefont {A.~N.}\ \bibnamefont {Antonov}}, \bibinfo
  {author} {\bibfnamefont {D.~N.}\ \bibnamefont {Kadrev}}, \bibinfo {author}
  {\bibfnamefont {P.}~\bibnamefont {Sarriguren}}, \ and\ \bibinfo {author}
  {\bibfnamefont {E.~M.}\ \bibnamefont {Guerra}},\ }\href@noop {} {\bibfield
  {journal} {\bibinfo  {journal} {arXiv: Nuclear Theory}\ } (\bibinfo {year}
  {2020})}\BibitemShut {NoStop}%
\bibitem [{\citenamefont {Gaidarov}\ \emph {et~al.}(2011)\citenamefont
  {Gaidarov}, \citenamefont {Antonov}, \citenamefont {Sarriguren},\ and\
  \citenamefont {Moya~de Guerra}}]{PhysRevC.84.034316}%
  \BibitemOpen
  \bibfield  {author} {\bibinfo {author} {\bibfnamefont {M.~K.}\ \bibnamefont
  {Gaidarov}}, \bibinfo {author} {\bibfnamefont {A.~N.}\ \bibnamefont
  {Antonov}}, \bibinfo {author} {\bibfnamefont {P.}~\bibnamefont {Sarriguren}},
  \ and\ \bibinfo {author} {\bibfnamefont {E.}~\bibnamefont {Moya~de Guerra}},\
  }\href {\doibase 10.1103/PhysRevC.84.034316} {\bibfield  {journal} {\bibinfo
  {journal} {Phys. Rev. C}\ }\textbf {\bibinfo {volume} {84}},\ \bibinfo
  {pages} {034316} (\bibinfo {year} {2011})}\BibitemShut {NoStop}%
\bibitem [{\citenamefont {Kaur}\ \emph {et~al.}(2020)\citenamefont {Kaur},
  \citenamefont {Quddus}, \citenamefont {Kumar}, \citenamefont {Bhuyan},\ and\
  \citenamefont {Patra}}]{Kaur2020OnTS}%
  \BibitemOpen
  \bibfield  {author} {\bibinfo {author} {\bibfnamefont {M.}~\bibnamefont
  {Kaur}}, \bibinfo {author} {\bibfnamefont {A.}~\bibnamefont {Quddus}},
  \bibinfo {author} {\bibfnamefont {A.}~\bibnamefont {Kumar}}, \bibinfo
  {author} {\bibfnamefont {M.}~\bibnamefont {Bhuyan}}, \ and\ \bibinfo {author}
  {\bibfnamefont {S.}~\bibnamefont {Patra}},\ }\href@noop {} {\bibfield
  {journal} {\bibinfo  {journal} {Journal of Physics G}\ }\textbf {\bibinfo
  {volume} {47}},\ \bibinfo {pages} {105102} (\bibinfo {year}
  {2020})}\BibitemShut {NoStop}%
\bibitem [{\citenamefont {Antonov}\ \emph {et~al.}(2018)\citenamefont
  {Antonov}, \citenamefont {Kadrev}, \citenamefont {Gaidarov}, \citenamefont
  {Sarriguren},\ and\ \citenamefont {Moya~de Guerra}}]{Antonov_2018}%
  \BibitemOpen
  \bibfield  {author} {\bibinfo {author} {\bibfnamefont {A.~N.}\ \bibnamefont
  {Antonov}}, \bibinfo {author} {\bibfnamefont {D.~N.}\ \bibnamefont {Kadrev}},
  \bibinfo {author} {\bibfnamefont {M.~K.}\ \bibnamefont {Gaidarov}}, \bibinfo
  {author} {\bibfnamefont {P.}~\bibnamefont {Sarriguren}}, \ and\ \bibinfo
  {author} {\bibfnamefont {E.}~\bibnamefont {Moya~de Guerra}},\ }\href
  {\doibase 10.1103/PhysRevC.98.054315} {\bibfield  {journal} {\bibinfo
  {journal} {Phys. Rev. C}\ }\textbf {\bibinfo {volume} {98}},\ \bibinfo
  {pages} {054315} (\bibinfo {year} {2018})}\BibitemShut {NoStop}%
\bibitem [{\citenamefont {Antonov}\ \emph {et~al.}(2016)\citenamefont
  {Antonov}, \citenamefont {Gaidarov}, \citenamefont {Sarriguren},\ and\
  \citenamefont {Moya~de Guerra}}]{Antonov_2016}%
  \BibitemOpen
  \bibfield  {author} {\bibinfo {author} {\bibfnamefont {A.~N.}\ \bibnamefont
  {Antonov}}, \bibinfo {author} {\bibfnamefont {M.~K.}\ \bibnamefont
  {Gaidarov}}, \bibinfo {author} {\bibfnamefont {P.}~\bibnamefont
  {Sarriguren}}, \ and\ \bibinfo {author} {\bibfnamefont {E.}~\bibnamefont
  {Moya~de Guerra}},\ }\href {\doibase 10.1103/PhysRevC.94.014319} {\bibfield
  {journal} {\bibinfo  {journal} {Phys. Rev. C}\ }\textbf {\bibinfo {volume}
  {94}},\ \bibinfo {pages} {014319} (\bibinfo {year} {2016})}\BibitemShut
  {NoStop}%
\bibitem [{\citenamefont {Gaidarov}\ \emph {et~al.}(2012)\citenamefont
  {Gaidarov}, \citenamefont {Antonov}, \citenamefont {Sarriguren},\ and\
  \citenamefont {de~Guerra}}]{PhysRevC.85.064319}%
  \BibitemOpen
  \bibfield  {author} {\bibinfo {author} {\bibfnamefont {M.~K.}\ \bibnamefont
  {Gaidarov}}, \bibinfo {author} {\bibfnamefont {A.~N.}\ \bibnamefont
  {Antonov}}, \bibinfo {author} {\bibfnamefont {P.}~\bibnamefont {Sarriguren}},
  \ and\ \bibinfo {author} {\bibfnamefont {E.~M.}\ \bibnamefont {de~Guerra}},\
  }\href {\doibase 10.1103/PhysRevC.85.064319} {\bibfield  {journal} {\bibinfo
  {journal} {Phys. Rev. C}\ }\textbf {\bibinfo {volume} {85}},\ \bibinfo
  {pages} {064319} (\bibinfo {year} {2012})}\BibitemShut {NoStop}%
\bibitem [{\citenamefont {Fetter}\ and\ \citenamefont
  {Walecka}(1971)}]{Fetter}%
  \BibitemOpen
  \bibfield  {author} {\bibinfo {author} {\bibfnamefont {A.~L.}\ \bibnamefont
  {Fetter}}\ and\ \bibinfo {author} {\bibfnamefont {J.~D.}\ \bibnamefont
  {Walecka}},\ }\href@noop {} {\emph {\bibinfo {title} {Quantum Theory of
  Many-Particle Systems}}}\ (\bibinfo  {publisher} {McGraw-Hill},\ \bibinfo
  {address} {Boston},\ \bibinfo {year} {1971})\BibitemShut {NoStop}%
\bibitem [{\citenamefont {Chen}\ \emph {et~al.}(2009)\citenamefont {Chen},
  \citenamefont {Cai}, \citenamefont {Ko}, \citenamefont {Li}, \citenamefont
  {Shen},\ and\ \citenamefont {Xu}}]{PhysRevC.80.014322}%
  \BibitemOpen
  \bibfield  {author} {\bibinfo {author} {\bibfnamefont {L.-W.}\ \bibnamefont
  {Chen}}, \bibinfo {author} {\bibfnamefont {B.-J.}\ \bibnamefont {Cai}},
  \bibinfo {author} {\bibfnamefont {C.~M.}\ \bibnamefont {Ko}}, \bibinfo
  {author} {\bibfnamefont {B.-A.}\ \bibnamefont {Li}}, \bibinfo {author}
  {\bibfnamefont {C.}~\bibnamefont {Shen}}, \ and\ \bibinfo {author}
  {\bibfnamefont {J.}~\bibnamefont {Xu}},\ }\href {\doibase
  10.1103/PhysRevC.80.014322} {\bibfield  {journal} {\bibinfo  {journal} {Phys.
  Rev. C}\ }\textbf {\bibinfo {volume} {80}},\ \bibinfo {pages} {014322}
  (\bibinfo {year} {2009})}\BibitemShut {NoStop}%
\bibitem [{\citenamefont {Chen}\ and\ \citenamefont
  {Piekarewicz}(2014)}]{PhysRevC.90.044305}%
  \BibitemOpen
  \bibfield  {author} {\bibinfo {author} {\bibfnamefont {W.-C.}\ \bibnamefont
  {Chen}}\ and\ \bibinfo {author} {\bibfnamefont {J.}~\bibnamefont
  {Piekarewicz}},\ }\href {\doibase 10.1103/PhysRevC.90.044305} {\bibfield
  {journal} {\bibinfo  {journal} {Phys. Rev. C}\ }\textbf {\bibinfo {volume}
  {90}},\ \bibinfo {pages} {044305} (\bibinfo {year} {2014})}\BibitemShut
  {NoStop}%
\bibitem [{\citenamefont {{Pacini}}(1965)}]{1965MmSAI36323P}%
  \BibitemOpen
  \bibfield  {author} {\bibinfo {author} {\bibfnamefont {F.}~\bibnamefont
  {{Pacini}}},\ }\href@noop {} {\bibfield  {journal} {\bibinfo  {journal}
  {memsai}\ }\textbf {\bibinfo {volume} {36}},\ \bibinfo {pages} {323}
  (\bibinfo {year} {1965})}\BibitemShut {NoStop}%
\bibitem [{\citenamefont {Oppenheimer}\ and\ \citenamefont
  {Volkoff}(1939)}]{PhysRev.55.374}%
  \BibitemOpen
  \bibfield  {author} {\bibinfo {author} {\bibfnamefont {J.~R.}\ \bibnamefont
  {Oppenheimer}}\ and\ \bibinfo {author} {\bibfnamefont {G.~M.}\ \bibnamefont
  {Volkoff}},\ }\href {\doibase 10.1103/PhysRev.55.374} {\bibfield  {journal}
  {\bibinfo  {journal} {Phys. Rev.}\ }\textbf {\bibinfo {volume} {55}},\
  \bibinfo {pages} {374} (\bibinfo {year} {1939})}\BibitemShut {NoStop}%
\bibitem [{\citenamefont {Tolman}(1939)}]{PhysRev.55.364}%
  \BibitemOpen
  \bibfield  {author} {\bibinfo {author} {\bibfnamefont {R.~C.}\ \bibnamefont
  {Tolman}},\ }\href {\doibase 10.1103/PhysRev.55.364} {\bibfield  {journal}
  {\bibinfo  {journal} {Phys. Rev.}\ }\textbf {\bibinfo {volume} {55}},\
  \bibinfo {pages} {364} (\bibinfo {year} {1939})}\BibitemShut {NoStop}%
\bibitem [{\citenamefont {{Baym}}\ \emph {et~al.}(1971)\citenamefont {{Baym}},
  \citenamefont {{Pethick}},\ and\ \citenamefont
  {{Sutherland}}}]{1971ApJ...170..299B}%
  \BibitemOpen
  \bibfield  {author} {\bibinfo {author} {\bibfnamefont {G.}~\bibnamefont
  {{Baym}}}, \bibinfo {author} {\bibfnamefont {C.}~\bibnamefont {{Pethick}}}, \
  and\ \bibinfo {author} {\bibfnamefont {P.}~\bibnamefont {{Sutherland}}},\
  }\href {\doibase 10.1086/151216} {\bibfield  {journal} {\bibinfo  {journal}
  {\apj}\ }\textbf {\bibinfo {volume} {170}},\ \bibinfo {pages} {299} (\bibinfo
  {year} {1971})}\BibitemShut {NoStop}%
\bibitem [{\citenamefont {Das}\ \emph {et~al.}(2020{\natexlab{b}})\citenamefont
  {Das}, \citenamefont {Kumar}, \citenamefont {Kumar}, \citenamefont {Biswal},
  \citenamefont {Nakatsukasa}, \citenamefont {Li},\ and\ \citenamefont
  {Patra}}]{10.1093/mnras/staa1435}%
  \BibitemOpen
  \bibfield  {author} {\bibinfo {author} {\bibfnamefont {H.~C.}\ \bibnamefont
  {Das}}, \bibinfo {author} {\bibfnamefont {A.}~\bibnamefont {Kumar}}, \bibinfo
  {author} {\bibfnamefont {B.}~\bibnamefont {Kumar}}, \bibinfo {author}
  {\bibfnamefont {S.~K.}\ \bibnamefont {Biswal}}, \bibinfo {author}
  {\bibfnamefont {T.}~\bibnamefont {Nakatsukasa}}, \bibinfo {author}
  {\bibfnamefont {A.}~\bibnamefont {Li}}, \ and\ \bibinfo {author}
  {\bibfnamefont {S.~K.}\ \bibnamefont {Patra}},\ }\href {\doibase
  10.1093/mnras/staa1435} {\bibfield  {journal} {\bibinfo  {journal} {Monthly
  Notices of the Royal Astronomical Society}\ }\textbf {\bibinfo {volume}
  {495}},\ \bibinfo {pages} {4893} (\bibinfo {year}
  {2020}{\natexlab{b}})}\BibitemShut {NoStop}%
\bibitem [{\citenamefont {Miller}\ \emph {et~al.}(2019)\citenamefont {Miller},
  \citenamefont {Lamb}, \citenamefont {Dittmann} \emph {et~al.}}]{Miller_2019}%
  \BibitemOpen
  \bibfield  {author} {\bibinfo {author} {\bibfnamefont {M.~C.}\ \bibnamefont
  {Miller}}, \bibinfo {author} {\bibfnamefont {F.~K.}\ \bibnamefont {Lamb}},
  \bibinfo {author} {\bibfnamefont {A.~J.}\ \bibnamefont {Dittmann}},  \emph
  {et~al.},\ }\href {\doibase 10.3847/2041-8213/ab50c5} {\bibfield  {journal}
  {\bibinfo  {journal} {APJ}\ }\textbf {\bibinfo {volume} {887}},\ \bibinfo
  {pages} {L24} (\bibinfo {year} {2019})}\BibitemShut {NoStop}%
\bibitem [{\citenamefont {Riley}\ \emph {et~al.}(2019)\citenamefont {Riley},
  \citenamefont {Watts}, \citenamefont {Bogdanov}, \citenamefont {Ray},
  \citenamefont {Ludlam}, \citenamefont {Guillot}, \citenamefont {Arzoumanian},
  \citenamefont {Baker}, \citenamefont {Bilous}, \citenamefont {Chakrabarty},
  \citenamefont {Gendreau}, \citenamefont {Harding}, \citenamefont {Ho},
  \citenamefont {Lattimer}, \citenamefont {Morsink},\ and\ \citenamefont
  {Strohmayer}}]{Riley_2019}%
  \BibitemOpen
  \bibfield  {author} {\bibinfo {author} {\bibfnamefont {T.~E.}\ \bibnamefont
  {Riley}}, \bibinfo {author} {\bibfnamefont {A.~L.}\ \bibnamefont {Watts}},
  \bibinfo {author} {\bibfnamefont {S.}~\bibnamefont {Bogdanov}}, \bibinfo
  {author} {\bibfnamefont {P.~S.}\ \bibnamefont {Ray}}, \bibinfo {author}
  {\bibfnamefont {R.~M.}\ \bibnamefont {Ludlam}}, \bibinfo {author}
  {\bibfnamefont {S.}~\bibnamefont {Guillot}}, \bibinfo {author} {\bibfnamefont
  {Z.}~\bibnamefont {Arzoumanian}}, \bibinfo {author} {\bibfnamefont {C.~L.}\
  \bibnamefont {Baker}}, \bibinfo {author} {\bibfnamefont {A.~V.}\ \bibnamefont
  {Bilous}}, \bibinfo {author} {\bibfnamefont {D.}~\bibnamefont {Chakrabarty}},
  \bibinfo {author} {\bibfnamefont {K.~C.}\ \bibnamefont {Gendreau}}, \bibinfo
  {author} {\bibfnamefont {A.~K.}\ \bibnamefont {Harding}}, \bibinfo {author}
  {\bibfnamefont {W.~C.~G.}\ \bibnamefont {Ho}}, \bibinfo {author}
  {\bibfnamefont {J.~M.}\ \bibnamefont {Lattimer}}, \bibinfo {author}
  {\bibfnamefont {S.~M.}\ \bibnamefont {Morsink}}, \ and\ \bibinfo {author}
  {\bibfnamefont {T.~E.}\ \bibnamefont {Strohmayer}},\ }\href {\doibase
  10.3847/2041-8213/ab481c} {\bibfield  {journal} {\bibinfo  {journal} {The
  Astrophysical Journal}\ }\textbf {\bibinfo {volume} {887}},\ \bibinfo {pages}
  {L21} (\bibinfo {year} {2019})}\BibitemShut {NoStop}%
\bibitem [{\citenamefont {Miller}\ \emph {et~al.}(2021)\citenamefont {Miller},
  \citenamefont {Lamb}, \citenamefont {Dittmann}, \citenamefont {Bogdanov},
  \citenamefont {Arzoumanian}, \citenamefont {Gendreau}, \citenamefont
  {Guillot}, \citenamefont {Ho}, \citenamefont {Lattimer}, \citenamefont
  {Loewenstein}, \citenamefont {Morsink}, \citenamefont {Ray}, \citenamefont
  {Wolff}, \citenamefont {Baker}, \citenamefont {Cazeau}, \citenamefont
  {Manthripragada}, \citenamefont {Markwardt}, \citenamefont {Okajima},
  \citenamefont {Pollard}, \citenamefont {Cognard}, \citenamefont {Cromartie},
  \citenamefont {Fonseca}, \citenamefont {Guillemot}, \citenamefont {Kerr},
  \citenamefont {Parthasarathy}, \citenamefont {Pennucci}, \citenamefont
  {Ransom},\ and\ \citenamefont {Stairs}}]{miller2021radius}%
  \BibitemOpen
  \bibfield  {author} {\bibinfo {author} {\bibfnamefont {M.~C.}\ \bibnamefont
  {Miller}}, \bibinfo {author} {\bibfnamefont {F.~K.}\ \bibnamefont {Lamb}},
  \bibinfo {author} {\bibfnamefont {A.~J.}\ \bibnamefont {Dittmann}}, \bibinfo
  {author} {\bibfnamefont {S.}~\bibnamefont {Bogdanov}}, \bibinfo {author}
  {\bibfnamefont {Z.}~\bibnamefont {Arzoumanian}}, \bibinfo {author}
  {\bibfnamefont {K.~C.}\ \bibnamefont {Gendreau}}, \bibinfo {author}
  {\bibfnamefont {S.}~\bibnamefont {Guillot}}, \bibinfo {author} {\bibfnamefont
  {W.~C.~G.}\ \bibnamefont {Ho}}, \bibinfo {author} {\bibfnamefont {J.~M.}\
  \bibnamefont {Lattimer}}, \bibinfo {author} {\bibfnamefont {M.}~\bibnamefont
  {Loewenstein}}, \bibinfo {author} {\bibfnamefont {S.~M.}\ \bibnamefont
  {Morsink}}, \bibinfo {author} {\bibfnamefont {P.~S.}\ \bibnamefont {Ray}},
  \bibinfo {author} {\bibfnamefont {M.~T.}\ \bibnamefont {Wolff}}, \bibinfo
  {author} {\bibfnamefont {C.~L.}\ \bibnamefont {Baker}}, \bibinfo {author}
  {\bibfnamefont {T.}~\bibnamefont {Cazeau}}, \bibinfo {author} {\bibfnamefont
  {S.}~\bibnamefont {Manthripragada}}, \bibinfo {author} {\bibfnamefont
  {C.~B.}\ \bibnamefont {Markwardt}}, \bibinfo {author} {\bibfnamefont
  {T.}~\bibnamefont {Okajima}}, \bibinfo {author} {\bibfnamefont
  {S.}~\bibnamefont {Pollard}}, \bibinfo {author} {\bibfnamefont
  {I.}~\bibnamefont {Cognard}}, \bibinfo {author} {\bibfnamefont {H.~T.}\
  \bibnamefont {Cromartie}}, \bibinfo {author} {\bibfnamefont {E.}~\bibnamefont
  {Fonseca}}, \bibinfo {author} {\bibfnamefont {L.}~\bibnamefont {Guillemot}},
  \bibinfo {author} {\bibfnamefont {M.}~\bibnamefont {Kerr}}, \bibinfo {author}
  {\bibfnamefont {A.}~\bibnamefont {Parthasarathy}}, \bibinfo {author}
  {\bibfnamefont {T.~T.}\ \bibnamefont {Pennucci}}, \bibinfo {author}
  {\bibfnamefont {S.}~\bibnamefont {Ransom}}, \ and\ \bibinfo {author}
  {\bibfnamefont {I.}~\bibnamefont {Stairs}},\ }\href@noop {} {\enquote
  {\bibinfo {title} {The radius of psr j0740+6620 from nicer and xmm-newton
  data},}\ } (\bibinfo {year} {2021}),\ \Eprint
  {http://arxiv.org/abs/2105.06979} {arXiv:2105.06979 [astro-ph.HE]}
  \BibitemShut {NoStop}%
\bibitem [{\citenamefont {Das}\ \emph {et~al.}(2021{\natexlab{b}})\citenamefont
  {Das}, \citenamefont {Kumar}, \citenamefont {Kumar}, \citenamefont {Biswal},\
  and\ \citenamefont {Patra}}]{Das_2021}%
  \BibitemOpen
  \bibfield  {author} {\bibinfo {author} {\bibfnamefont {H.~C.}\ \bibnamefont
  {Das}}, \bibinfo {author} {\bibfnamefont {A.}~\bibnamefont {Kumar}}, \bibinfo
  {author} {\bibfnamefont {B.}~\bibnamefont {Kumar}}, \bibinfo {author}
  {\bibfnamefont {S.~K.}\ \bibnamefont {Biswal}}, \ and\ \bibinfo {author}
  {\bibfnamefont {S.~K.}\ \bibnamefont {Patra}},\ }\href {\doibase
  10.1088/1475-7516/2021/01/007} {\bibfield  {journal} {\bibinfo  {journal}
  {Journal of Cosmology and Astroparticle Physics}\ }\textbf {\bibinfo {volume}
  {2021}},\ \bibinfo {pages} {007} (\bibinfo {year}
  {2021}{\natexlab{b}})}\BibitemShut {NoStop}%
\bibitem [{\citenamefont {Arzoumanian}\ \emph {et~al.}(2018)\citenamefont
  {Arzoumanian}, \citenamefont {Brazier}, \citenamefont {Burke-Spolaor},
  \citenamefont {Chamberlin}, \citenamefont {Chatterjee}, \citenamefont
  {Christy}, \citenamefont {Cordes}, \citenamefont {Cornish}, \citenamefont
  {Crawford}, \citenamefont {Cromartie}, \citenamefont {Crowter}, \citenamefont
  {DeCesar}, \citenamefont {Demorest}, \citenamefont {Dolch}, \citenamefont
  {Ellis}, \citenamefont {Ferdman}, \citenamefont {Ferrara}, \citenamefont
  {Fonseca}, \citenamefont {Garver-Daniels}, \citenamefont {Gentile},
  \citenamefont {Halmrast}, \citenamefont {Huerta}, \citenamefont {Jenet},
  \citenamefont {Jessup}, \citenamefont {Jones}, \citenamefont {Jones},
  \citenamefont {Kaplan}, \citenamefont {Lam}, \citenamefont {Lazio},
  \citenamefont {Levin}, \citenamefont {Lommen}, \citenamefont {Lorimer},
  \citenamefont {Luo}, \citenamefont {Lynch}, \citenamefont {Madison},
  \citenamefont {Matthews}, \citenamefont {McLaughlin}, \citenamefont
  {McWilliams}, \citenamefont {Mingarelli}, \citenamefont {Ng}, \citenamefont
  {Nice}, \citenamefont {Pennucci}, \citenamefont {Ransom}, \citenamefont
  {Ray}, \citenamefont {Siemens}, \citenamefont {Simon}, \citenamefont
  {Spiewak}, \citenamefont {Stairs}, \citenamefont {Stinebring}, \citenamefont
  {Stovall}, \citenamefont {Swiggum}, \citenamefont {Taylor}, \citenamefont
  {Vallisneri}, \citenamefont {van Haasteren}, \citenamefont {Vigeland},\ and\
  \citenamefont {and}}]{Arzoumanian_2018}%
  \BibitemOpen
  \bibfield  {author} {\bibinfo {author} {\bibfnamefont {Z.}~\bibnamefont
  {Arzoumanian}}, \bibinfo {author} {\bibfnamefont {A.}~\bibnamefont
  {Brazier}}, \bibinfo {author} {\bibfnamefont {S.}~\bibnamefont
  {Burke-Spolaor}}, \bibinfo {author} {\bibfnamefont {S.}~\bibnamefont
  {Chamberlin}}, \bibinfo {author} {\bibfnamefont {S.}~\bibnamefont
  {Chatterjee}}, \bibinfo {author} {\bibfnamefont {B.}~\bibnamefont {Christy}},
  \bibinfo {author} {\bibfnamefont {J.~M.}\ \bibnamefont {Cordes}}, \bibinfo
  {author} {\bibfnamefont {N.~J.}\ \bibnamefont {Cornish}}, \bibinfo {author}
  {\bibfnamefont {F.}~\bibnamefont {Crawford}}, \bibinfo {author}
  {\bibfnamefont {H.~T.}\ \bibnamefont {Cromartie}}, \bibinfo {author}
  {\bibfnamefont {K.}~\bibnamefont {Crowter}}, \bibinfo {author} {\bibfnamefont
  {M.~E.}\ \bibnamefont {DeCesar}}, \bibinfo {author} {\bibfnamefont {P.~B.}\
  \bibnamefont {Demorest}}, \bibinfo {author} {\bibfnamefont {T.}~\bibnamefont
  {Dolch}}, \bibinfo {author} {\bibfnamefont {J.~A.}\ \bibnamefont {Ellis}},
  \bibinfo {author} {\bibfnamefont {R.~D.}\ \bibnamefont {Ferdman}}, \bibinfo
  {author} {\bibfnamefont {E.~C.}\ \bibnamefont {Ferrara}}, \bibinfo {author}
  {\bibfnamefont {E.}~\bibnamefont {Fonseca}}, \bibinfo {author} {\bibfnamefont
  {N.}~\bibnamefont {Garver-Daniels}}, \bibinfo {author} {\bibfnamefont
  {P.~A.}\ \bibnamefont {Gentile}}, \bibinfo {author} {\bibfnamefont
  {D.}~\bibnamefont {Halmrast}}, \bibinfo {author} {\bibfnamefont {E.~A.}\
  \bibnamefont {Huerta}}, \bibinfo {author} {\bibfnamefont {F.~A.}\
  \bibnamefont {Jenet}}, \bibinfo {author} {\bibfnamefont {C.}~\bibnamefont
  {Jessup}}, \bibinfo {author} {\bibfnamefont {G.}~\bibnamefont {Jones}},
  \bibinfo {author} {\bibfnamefont {M.~L.}\ \bibnamefont {Jones}}, \bibinfo
  {author} {\bibfnamefont {D.~L.}\ \bibnamefont {Kaplan}}, \bibinfo {author}
  {\bibfnamefont {M.~T.}\ \bibnamefont {Lam}}, \bibinfo {author} {\bibfnamefont
  {T.~J.~W.}\ \bibnamefont {Lazio}}, \bibinfo {author} {\bibfnamefont
  {L.}~\bibnamefont {Levin}}, \bibinfo {author} {\bibfnamefont
  {A.}~\bibnamefont {Lommen}}, \bibinfo {author} {\bibfnamefont {D.~R.}\
  \bibnamefont {Lorimer}}, \bibinfo {author} {\bibfnamefont {J.}~\bibnamefont
  {Luo}}, \bibinfo {author} {\bibfnamefont {R.~S.}\ \bibnamefont {Lynch}},
  \bibinfo {author} {\bibfnamefont {D.}~\bibnamefont {Madison}}, \bibinfo
  {author} {\bibfnamefont {A.~M.}\ \bibnamefont {Matthews}}, \bibinfo {author}
  {\bibfnamefont {M.~A.}\ \bibnamefont {McLaughlin}}, \bibinfo {author}
  {\bibfnamefont {S.~T.}\ \bibnamefont {McWilliams}}, \bibinfo {author}
  {\bibfnamefont {C.}~\bibnamefont {Mingarelli}}, \bibinfo {author}
  {\bibfnamefont {C.}~\bibnamefont {Ng}}, \bibinfo {author} {\bibfnamefont
  {D.~J.}\ \bibnamefont {Nice}}, \bibinfo {author} {\bibfnamefont {T.~T.}\
  \bibnamefont {Pennucci}}, \bibinfo {author} {\bibfnamefont {S.~M.}\
  \bibnamefont {Ransom}}, \bibinfo {author} {\bibfnamefont {P.~S.}\
  \bibnamefont {Ray}}, \bibinfo {author} {\bibfnamefont {X.}~\bibnamefont
  {Siemens}}, \bibinfo {author} {\bibfnamefont {J.}~\bibnamefont {Simon}},
  \bibinfo {author} {\bibfnamefont {R.}~\bibnamefont {Spiewak}}, \bibinfo
  {author} {\bibfnamefont {I.~H.}\ \bibnamefont {Stairs}}, \bibinfo {author}
  {\bibfnamefont {D.~R.}\ \bibnamefont {Stinebring}}, \bibinfo {author}
  {\bibfnamefont {K.}~\bibnamefont {Stovall}}, \bibinfo {author} {\bibfnamefont
  {J.~K.}\ \bibnamefont {Swiggum}}, \bibinfo {author} {\bibfnamefont {S.~R.}\
  \bibnamefont {Taylor}}, \bibinfo {author} {\bibfnamefont {M.}~\bibnamefont
  {Vallisneri}}, \bibinfo {author} {\bibfnamefont {R.}~\bibnamefont {van
  Haasteren}}, \bibinfo {author} {\bibfnamefont {S.~J.}\ \bibnamefont
  {Vigeland}}, \ and\ \bibinfo {author} {\bibfnamefont {W.~Z.}\ \bibnamefont
  {and}},\ }\href {\doibase 10.3847/1538-4365/aab5b0} {\bibfield  {journal}
  {\bibinfo  {journal} {The Astrophysical Journal Supplement Series}\ }\textbf
  {\bibinfo {volume} {235}},\ \bibinfo {pages} {37} (\bibinfo {year}
  {2018})}\BibitemShut {NoStop}%
\bibitem [{\citenamefont {Antoniadis}\ \emph {et~al.}(2013)\citenamefont
  {Antoniadis}, \citenamefont {Freire}, \citenamefont {Wex}, \citenamefont
  {Tauris}, \citenamefont {Lynch}, \citenamefont {van Kerkwijk}, \citenamefont
  {Kramer}, \citenamefont {Bassa}, \citenamefont {Dhillon}, \citenamefont
  {Driebe}, \citenamefont {Hessels}, \citenamefont {Kaspi}, \citenamefont
  {Kondratiev}, \citenamefont {Langer}, \citenamefont {Marsh}, \citenamefont
  {McLaughlin}, \citenamefont {Pennucci}, \citenamefont {Ransom}, \citenamefont
  {Stairs}, \citenamefont {van Leeuwen}, \citenamefont {Verbiest},\ and\
  \citenamefont {Whelan}}]{Antoniadis_2013}%
  \BibitemOpen
  \bibfield  {author} {\bibinfo {author} {\bibfnamefont {J.}~\bibnamefont
  {Antoniadis}}, \bibinfo {author} {\bibfnamefont {P.~C.~C.}\ \bibnamefont
  {Freire}}, \bibinfo {author} {\bibfnamefont {N.}~\bibnamefont {Wex}},
  \bibinfo {author} {\bibfnamefont {T.~M.}\ \bibnamefont {Tauris}}, \bibinfo
  {author} {\bibfnamefont {R.~S.}\ \bibnamefont {Lynch}}, \bibinfo {author}
  {\bibfnamefont {M.~H.}\ \bibnamefont {van Kerkwijk}}, \bibinfo {author}
  {\bibfnamefont {M.}~\bibnamefont {Kramer}}, \bibinfo {author} {\bibfnamefont
  {C.}~\bibnamefont {Bassa}}, \bibinfo {author} {\bibfnamefont {V.~S.}\
  \bibnamefont {Dhillon}}, \bibinfo {author} {\bibfnamefont {T.}~\bibnamefont
  {Driebe}}, \bibinfo {author} {\bibfnamefont {J.~W.~T.}\ \bibnamefont
  {Hessels}}, \bibinfo {author} {\bibfnamefont {V.~M.}\ \bibnamefont {Kaspi}},
  \bibinfo {author} {\bibfnamefont {V.~I.}\ \bibnamefont {Kondratiev}},
  \bibinfo {author} {\bibfnamefont {N.}~\bibnamefont {Langer}}, \bibinfo
  {author} {\bibfnamefont {T.~R.}\ \bibnamefont {Marsh}}, \bibinfo {author}
  {\bibfnamefont {M.~A.}\ \bibnamefont {McLaughlin}}, \bibinfo {author}
  {\bibfnamefont {T.~T.}\ \bibnamefont {Pennucci}}, \bibinfo {author}
  {\bibfnamefont {S.~M.}\ \bibnamefont {Ransom}}, \bibinfo {author}
  {\bibfnamefont {I.~H.}\ \bibnamefont {Stairs}}, \bibinfo {author}
  {\bibfnamefont {J.}~\bibnamefont {van Leeuwen}}, \bibinfo {author}
  {\bibfnamefont {J.~P.~W.}\ \bibnamefont {Verbiest}}, \ and\ \bibinfo {author}
  {\bibfnamefont {D.~G.}\ \bibnamefont {Whelan}},\ }\href {\doibase
  10.1126/science.1233232} {\bibfield  {journal} {\bibinfo  {journal}
  {Science}\ }\textbf {\bibinfo {volume} {340}},\ \bibinfo {pages} {1233232}
  (\bibinfo {year} {2013})}\BibitemShut {NoStop}%
\bibitem [{\citenamefont {Cromartie}\ \emph {et~al.}(2019)\citenamefont
  {Cromartie}, \citenamefont {Fonseca}, \citenamefont {Ransom}, \citenamefont
  {Demorest}, \citenamefont {Arzoumanian}, \citenamefont {Blumer},
  \citenamefont {Brook}, \citenamefont {DeCesar}, \citenamefont {Dolch},
  \citenamefont {Ellis},\ and\ \citenamefont {et~al.}}]{Cromartie_2019}%
  \BibitemOpen
  \bibfield  {author} {\bibinfo {author} {\bibfnamefont {H.~T.}\ \bibnamefont
  {Cromartie}}, \bibinfo {author} {\bibfnamefont {E.}~\bibnamefont {Fonseca}},
  \bibinfo {author} {\bibfnamefont {S.~M.}\ \bibnamefont {Ransom}}, \bibinfo
  {author} {\bibfnamefont {P.~B.}\ \bibnamefont {Demorest}}, \bibinfo {author}
  {\bibfnamefont {Z.}~\bibnamefont {Arzoumanian}}, \bibinfo {author}
  {\bibfnamefont {H.}~\bibnamefont {Blumer}}, \bibinfo {author} {\bibfnamefont
  {P.~R.}\ \bibnamefont {Brook}}, \bibinfo {author} {\bibfnamefont {M.~E.}\
  \bibnamefont {DeCesar}}, \bibinfo {author} {\bibfnamefont {T.}~\bibnamefont
  {Dolch}}, \bibinfo {author} {\bibfnamefont {J.~A.}\ \bibnamefont {Ellis}}, \
  and\ \bibinfo {author} {\bibnamefont {et~al.}},\ }\href {\doibase
  10.1038/s41550-019-0880-2} {\bibfield  {journal} {\bibinfo  {journal} {Nature
  Astronomy}\ }\textbf {\bibinfo {volume} {4}},\ \bibinfo {pages} {72–76}
  (\bibinfo {year} {2019})}\BibitemShut {NoStop}%
\bibitem [{\citenamefont {Fonseca}\ \emph {et~al.}(2021)\citenamefont
  {Fonseca}, \citenamefont {Cromartie}, \citenamefont {Pennucci}, \citenamefont
  {Ray}, \citenamefont {Kirichenko}, \citenamefont {Ransom}, \citenamefont
  {Demorest}, \citenamefont {Stairs}, \citenamefont {Arzoumanian},
  \citenamefont {Guillemot},\ and\ \citenamefont {et~al.}}]{Fonseca_2021}%
  \BibitemOpen
  \bibfield  {author} {\bibinfo {author} {\bibfnamefont {E.}~\bibnamefont
  {Fonseca}}, \bibinfo {author} {\bibfnamefont {H.~T.}\ \bibnamefont
  {Cromartie}}, \bibinfo {author} {\bibfnamefont {T.~T.}\ \bibnamefont
  {Pennucci}}, \bibinfo {author} {\bibfnamefont {P.~S.}\ \bibnamefont {Ray}},
  \bibinfo {author} {\bibfnamefont {A.~Y.}\ \bibnamefont {Kirichenko}},
  \bibinfo {author} {\bibfnamefont {S.~M.}\ \bibnamefont {Ransom}}, \bibinfo
  {author} {\bibfnamefont {P.~B.}\ \bibnamefont {Demorest}}, \bibinfo {author}
  {\bibfnamefont {I.~H.}\ \bibnamefont {Stairs}}, \bibinfo {author}
  {\bibfnamefont {Z.}~\bibnamefont {Arzoumanian}}, \bibinfo {author}
  {\bibfnamefont {L.}~\bibnamefont {Guillemot}}, \ and\ \bibinfo {author}
  {\bibnamefont {et~al.}},\ }\href {\doibase 10.3847/2041-8213/ac03b8}
  {\bibfield  {journal} {\bibinfo  {journal} {The Astrophysical Journal
  Letters}\ }\textbf {\bibinfo {volume} {915}},\ \bibinfo {pages} {L12}
  (\bibinfo {year} {2021})}\BibitemShut {NoStop}%
\bibitem [{\citenamefont {Kumar}\ \emph
  {et~al.}(2018{\natexlab{b}})\citenamefont {Kumar}, \citenamefont {Patra},\
  and\ \citenamefont {Agrawal}}]{PhysRevC.97.045806}%
  \BibitemOpen
  \bibfield  {author} {\bibinfo {author} {\bibfnamefont {B.}~\bibnamefont
  {Kumar}}, \bibinfo {author} {\bibfnamefont {S.~K.}\ \bibnamefont {Patra}}, \
  and\ \bibinfo {author} {\bibfnamefont {B.~K.}\ \bibnamefont {Agrawal}},\
  }\href {\doibase 10.1103/PhysRevC.97.045806} {\bibfield  {journal} {\bibinfo
  {journal} {Phys. Rev. C}\ }\textbf {\bibinfo {volume} {97}},\ \bibinfo
  {pages} {045806} (\bibinfo {year} {2018}{\natexlab{b}})}\BibitemShut
  {NoStop}%
\bibitem [{\citenamefont {Das}\ \emph {et~al.}(2021{\natexlab{c}})\citenamefont
  {Das}, \citenamefont {Kumar},\ and\ \citenamefont {Patra}}]{das2021effects}%
  \BibitemOpen
  \bibfield  {author} {\bibinfo {author} {\bibfnamefont {H.~C.}\ \bibnamefont
  {Das}}, \bibinfo {author} {\bibfnamefont {A.}~\bibnamefont {Kumar}}, \ and\
  \bibinfo {author} {\bibfnamefont {S.~K.}\ \bibnamefont {Patra}},\ }\href@noop
  {} {} (\bibinfo {year} {2021}{\natexlab{c}}),\ \Eprint
  {http://arxiv.org/abs/2104.01815} {arXiv:2104.01815 [astro-ph.HE]}
  \BibitemShut {NoStop}%
\bibitem [{\citenamefont {Glendenning}(1997)}]{Glendenning:1997wn}%
  \BibitemOpen
  \bibfield  {author} {\bibinfo {author} {\bibfnamefont {N.~K.}\ \bibnamefont
  {Glendenning}},\ }\href@noop {} {\emph {\bibinfo {title} {{Compact stars:
  Nuclear physics, particle physics, and general relativity}}}}\ (\bibinfo
  {year} {1997})\BibitemShut {NoStop}%
\bibitem [{\citenamefont {Patra}\ \emph
  {et~al.}(2002{\natexlab{b}})\citenamefont {Patra}, \citenamefont {Raj},
  \citenamefont {Mehta},\ and\ \citenamefont {Gupta}}]{PhysRevC.65.054323}%
  \BibitemOpen
  \bibfield  {author} {\bibinfo {author} {\bibfnamefont {S.~K.}\ \bibnamefont
  {Patra}}, \bibinfo {author} {\bibfnamefont {B.~K.}\ \bibnamefont {Raj}},
  \bibinfo {author} {\bibfnamefont {M.~S.}\ \bibnamefont {Mehta}}, \ and\
  \bibinfo {author} {\bibfnamefont {R.~K.}\ \bibnamefont {Gupta}},\ }\href
  {\doibase 10.1103/PhysRevC.65.054323} {\bibfield  {journal} {\bibinfo
  {journal} {Phys. Rev. C}\ }\textbf {\bibinfo {volume} {65}},\ \bibinfo
  {pages} {054323} (\bibinfo {year} {2002}{\natexlab{b}})}\BibitemShut
  {NoStop}%
\bibitem [{\citenamefont {Patra}\ \emph {et~al.}(2001)\citenamefont {Patra},
  \citenamefont {Centelles}, \citenamefont {Viñas},\ and\ \citenamefont {{Del
  Estal}}}]{PATRA200167}%
  \BibitemOpen
  \bibfield  {author} {\bibinfo {author} {\bibfnamefont {S.}~\bibnamefont
  {Patra}}, \bibinfo {author} {\bibfnamefont {M.}~\bibnamefont {Centelles}},
  \bibinfo {author} {\bibfnamefont {X.}~\bibnamefont {Viñas}}, \ and\ \bibinfo
  {author} {\bibfnamefont {M.}~\bibnamefont {{Del Estal}}},\ }\href {\doibase
  https://doi.org/10.1016/S0370-2693(01)01328-4} {\bibfield  {journal}
  {\bibinfo  {journal} {Physics Letters B}\ }\textbf {\bibinfo {volume}
  {523}},\ \bibinfo {pages} {67} (\bibinfo {year} {2001})}\BibitemShut
  {NoStop}%
\bibitem [{\citenamefont {Patra}\ \emph
  {et~al.}(2002{\natexlab{c}})\citenamefont {Patra}, \citenamefont {Viñas},
  \citenamefont {Centelles},\ and\ \citenamefont {{Del Estal}}}]{PATRA2002240}%
  \BibitemOpen
  \bibfield  {author} {\bibinfo {author} {\bibfnamefont {S.}~\bibnamefont
  {Patra}}, \bibinfo {author} {\bibfnamefont {X.}~\bibnamefont {Viñas}},
  \bibinfo {author} {\bibfnamefont {M.}~\bibnamefont {Centelles}}, \ and\
  \bibinfo {author} {\bibfnamefont {M.}~\bibnamefont {{Del Estal}}},\ }\href
  {\doibase https://doi.org/10.1016/S0375-9474(01)01531-7} {\bibfield
  {journal} {\bibinfo  {journal} {Nuclear Physics A}\ }\textbf {\bibinfo
  {volume} {703}},\ \bibinfo {pages} {240} (\bibinfo {year}
  {2002}{\natexlab{c}})}\BibitemShut {NoStop}%
\bibitem [{\citenamefont {Pattnaik}\ \emph {et~al.}(2021)\citenamefont
  {Pattnaik}, \citenamefont {Kumar}, \citenamefont {Das}, \citenamefont
  {Bhuyan},\ and\ \citenamefont {Patra}}]{Pattnaik}%
  \BibitemOpen
  \bibfield  {author} {\bibinfo {author} {\bibfnamefont {J.~A.}\ \bibnamefont
  {Pattnaik}}, \bibinfo {author} {\bibfnamefont {A.}~\bibnamefont {Kumar}},
  \bibinfo {author} {\bibfnamefont {H.~C.}\ \bibnamefont {Das}}, \bibinfo
  {author} {\bibfnamefont {M.}~\bibnamefont {Bhuyan}}, \ and\ \bibinfo {author}
  {\bibfnamefont {S.~K.}\ \bibnamefont {Patra}},\ }\href@noop {} {\bibfield
  {journal} {\bibinfo  {journal} {in preparartion}\ } (\bibinfo {year}
  {2021})}\BibitemShut {NoStop}%
\bibitem [{\citenamefont {{Lattimer}}(2015)}]{2015AIPC.1645...61L}%
  \BibitemOpen
  \bibfield  {author} {\bibinfo {author} {\bibfnamefont {J.~M.}\ \bibnamefont
  {{Lattimer}}},\ }in\ \href {\doibase 10.1063/1.4909560} {\emph {\bibinfo
  {booktitle} {Exotic Nuclei and Nuclear/Particle Astrophysics (V) From Nuclei
  to Stars: Carpathian Summer School of Physics 2014}}},\ \bibinfo {series}
  {American Institute of Physics Conference Series}, Vol.\ \bibinfo {volume}
  {1645}\ (\bibinfo {year} {2015})\ pp.\ \bibinfo {pages} {61--78}\BibitemShut
  {NoStop}%
\bibitem [{\citenamefont {Lattimer}\ and\ \citenamefont
  {Lim}(2013)}]{Lattimer_2013}%
  \BibitemOpen
  \bibfield  {author} {\bibinfo {author} {\bibfnamefont {J.~M.}\ \bibnamefont
  {Lattimer}}\ and\ \bibinfo {author} {\bibfnamefont {Y.}~\bibnamefont {Lim}},\
  }\href {\doibase 10.1088/0004-637x/771/1/51} {\bibfield  {journal} {\bibinfo
  {journal} {The Astrophysical Journal}\ }\textbf {\bibinfo {volume} {771}},\
  \bibinfo {pages} {51} (\bibinfo {year} {2013})}\BibitemShut {NoStop}%
\bibitem [{\citenamefont {Chen}\ \emph {et~al.}(2007)\citenamefont {Chen},
  \citenamefont {Ko},\ and\ \citenamefont {Li}}]{PhysRevC.76.054316}%
  \BibitemOpen
  \bibfield  {author} {\bibinfo {author} {\bibfnamefont {L.-W.}\ \bibnamefont
  {Chen}}, \bibinfo {author} {\bibfnamefont {C.~M.}\ \bibnamefont {Ko}}, \ and\
  \bibinfo {author} {\bibfnamefont {B.-A.}\ \bibnamefont {Li}},\ }\href
  {\doibase 10.1103/PhysRevC.76.054316} {\bibfield  {journal} {\bibinfo
  {journal} {Phys. Rev. C}\ }\textbf {\bibinfo {volume} {76}},\ \bibinfo
  {pages} {054316} (\bibinfo {year} {2007})}\BibitemShut {NoStop}%
\bibitem [{\citenamefont {Centelles}\ \emph {et~al.}(2009)\citenamefont
  {Centelles}, \citenamefont {Roca-Maza}, \citenamefont {Vi\~nas},\ and\
  \citenamefont {Warda}}]{PhysRevLett.102.122502}%
  \BibitemOpen
  \bibfield  {author} {\bibinfo {author} {\bibfnamefont {M.}~\bibnamefont
  {Centelles}}, \bibinfo {author} {\bibfnamefont {X.}~\bibnamefont
  {Roca-Maza}}, \bibinfo {author} {\bibfnamefont {X.}~\bibnamefont {Vi\~nas}},
  \ and\ \bibinfo {author} {\bibfnamefont {M.}~\bibnamefont {Warda}},\ }\href
  {\doibase 10.1103/PhysRevLett.102.122502} {\bibfield  {journal} {\bibinfo
  {journal} {Phys. Rev. Lett.}\ }\textbf {\bibinfo {volume} {102}},\ \bibinfo
  {pages} {122502} (\bibinfo {year} {2009})}\BibitemShut {NoStop}%
\end{thebibliography}%
\bibliographystyle{apsrev4-1}
\end{document}